\documentclass[prb,twocolumn,showpacs,preprintnumbers,amsmath,amssymb,floatfix]
{revtex4}

\usepackage{soul}
        \setul{0.5ex}{}
\usepackage{graphicx}
\usepackage[center]{subfigure}
\usepackage{color}
\usepackage[usenames,dvipsnames,svgnames,table]{xcolor}
\usepackage{rotating}
\usepackage{placeins}
\newcommand \be {\begin{equation}}
\newcommand \ee {\end{equation}}
\newcommand \ben {\begin{eqnarray}}
\newcommand \een {\end{eqnarray}}

\newcommand \liqr {\rho_{\ell}}
\newcommand \sfe {\gamma_{\rm ISF}}

\begin{document}

\title{
Defect stability in phase-field crystal models:
Stacking faults and partial dislocations
}

\author{Joel Berry$^{1,3}$, 
Nikolas Provatas$^{1,2}$, 
J\"org Rottler$^3$, and 
Chad W.\ Sinclair$^4$
}

\affiliation{$^1$Department of Materials Science and Engineering,
McMaster University, 1280 Main Street West, Hamilton, Ontario, L8S 4L7, Canada}
\affiliation{$^2$Physics Department, McGill University,
3600 rue University, Montr\'eal, Qu\'ebec, H3A 2T8, Canada}
\affiliation{$^3$
Department of Physics and Astronomy, The University of British Columbia, 
6224 Agricultural Road, Vancouver, British Columbia, V6T 1Z1, Canada}
\affiliation{$^4$
Department of Materials Engineering, The University of British Columbia, 
309-6350 Stores Road, Vancouver, British Columbia, V6T 1Z4, Canada}

\date{\today}

\begin{abstract}
The primary factors controlling defect stability in phase-field
crystal (PFC) models are examined, with
illustrative examples involving several existing variations of the model.
Guidelines are presented for constructing models with stable
defect structures that maintain high numerical efficiency.
The general framework combines both long-range elastic fields and 
basic features of atomic-level core structures, 
with defect dynamics operable over diffusive time scales.
Fundamental elements of the resulting defect physics
are characterized for the case of fcc crystals.
Stacking faults and split Shockley partial dislocations
are stabilized for the first time within 
the PFC formalism, and
various properties of associated defect structures are characterized.
These include the dissociation width of perfect edge and screw dislocations,
the effect of applied stresses on dissociation, 
Peierls strains for glide, and dynamic contraction of
gliding pairs of partials.
Our results in general are shown to compare favorably with
continuum elastic theories and experimental findings.
\end{abstract}

\maketitle

\section{Introduction}
\label{sec:intro}
Structural kinetics in crystalline solids are driven heterogeneously 
at the atomic level by
localized defects, which in turn drive mesoscopic and macroscopic 
phenomena such as
structural phase transformations, fracture, and other forms of plastic flow.
A complete description of such processes therefore
requires a multiscale approach.
Existing modeling methods typically operate either exclusively on atomic
scales or on meso- and macroscopic scales.
Phase-field crystal models on the other hand provide a framework that
combines atomic length scales and mesoscopic/diffusive time scales
\cite{pfc02,pfc04,pfcdft07},
with the potential to reach mesoscopic lengths through systematic
multi-scale expansion methods
\cite{pfcRGnigel06a,pfcadmesh07,
pfcRGcoexist10,pfcbinaryamp10,pfcRGkarma10}.

The PFC approach naturally incorporates elasticity, plasticity,
and effects of local crystal orientation into a relatively simple atomic-level
continuum theory \cite{pfc02,pfc04}.
The literature published to date demonstrates
that such a formulation, with only a few minimal ingredients,
gives rise to a broad range of physics associated with
diffusive nonequilibrium processes in liquid and solid systems.
One of its strengths in terms of describing periodic systems is
the wealth of inherent defect structures that automatically emerge from
the basic free energy functional.
This permits the study of crystalline defects within a description that 
captures both
long-range elastic fields and basic features of atomic-level 
core structures.
In addition to fundamental, local defect properties, the
role of these defects in dynamic materials phenomena that operate over
long, diffusive time scales can be examined. Therefore,
defect stability, the core structure of stable defect configurations, 
and the dynamics and interactions of various defect structures within the PFC 
description all become central issues as the method is advanced further
into solid-state phenomena.
However, in light of the approximations inherent to the approach, 
its limitations need to be adequately realized and understood as well.

The basic properties of perfect PFC crystals are relatively well understood,
but the various classes and types of defects relevant
to each given lattice symmetry
require further study if material-specific applications are to be pursued. 
Perfect dislocations and simple
grain boundaries in 2D triangular and 3D bcc crystals
have been examined and
are known to be topologically stable under typical conditions of
small local strain and low thermodynamic driving force
\cite{pfc04,pfcdisloc06,
mpfc,mpfc09,pfcdeformscheme09,pfcavalanches10,pfcdft07,pfcbcc09,
pfccurved10,pfcfccGB10,pfcpremelt08,pfcpremeltkarma08,pfchotgb11}.
Certain other defect structures, 
notably stacking faults in close-packed 3D crystals, 
have inherently lower topological stability, and in such cases the proper
balance between crystal stability and defect stability 
in the model formulation becomes more restrictive and difficult to achieve.
It is also not necessarily clear which of the now many versions of PFC 
are best suited for describing defect-mediated processes.

The aims of the first part of this article are to characterize the
nature of defect stability in PFC models in terms of a few
general model features, and to examine how various versions
compare in this respect. 
We will show that fine control of crystal structure
often leads to reduced defect stability. A balance must therefore be
found that sufficiently 
favors
crystalline order yet does not
destabilize relatively unprotected defect structures such as planar faults. 
A few ways to achieve this balance
are identified and compared critically.
The most efficient of these approaches is then, in the second part
of this article, applied to
the case of fcc crystals, in which the stability of stacking faults
plays a central role. A range of defect properties and behaviors
consistent with theory and/or experiment are shown to naturally emerge
from what is still a very simple set of equations. 
Potential implications for atomistic studies of plastic deformation 
involving slow, diffusive processes are discussed, and initial results
concerning a few processes of interest are described. More detailed findings
relative to these will be outlined further in a future publication.

PFC models can be viewed as simplified versions of 
classical density functional theory (CDFT)
\cite{pfcdft07,singh91,oxtoby02}.
The Ramakrishnan-Yussouff (RY) CDFT \cite{ry79}
provides a useful reference point, with a free energy functional
given by an expansion around the liquid state correlation functions,
\begin{eqnarray}
\frac{F}{k_B T}&=&\int d\vec{r}~\left[ \rho(\vec{r}) \ln
\left( \rho(\vec{r})/{\liqr}\right)
-\delta \rho(\vec{r}) \right]- \nonumber \\
& &\frac{1}{2}
\int\int d\vec{r}~d\vec{r}_2~\delta \rho(\vec{r}) C_2(\vec{r},
\vec{r}_2) \delta \rho(\vec{r}_2)+\cdots
\label{dftfree}
\end{eqnarray}
where $\rho(\vec{r})$ is the atomic number density field,
$\liqr$ is a constant reference density,
$\delta \rho(\vec{r})=\rho(\vec{r})-\liqr$, and
$C_2(\vec{r},\vec{r}_2)=C_2(|\vec{r}-\vec{r}_2|)$,
is the two-point direct correlation function of the fluid, assumed isotropic.

A general PFC-type functional can be derived
from Eq.\ (\ref{dftfree}), truncated beyond $C_2(|\vec{r}-\vec{r}_2|)$,
as described in Ref.\ \onlinecite{pfcdft07}.
In terms of the rescaled atomic density field 
$n(\vec{r})=\rho(\vec{r})/\liqr-1$, 
an expansion of the logarithm in Eq.\ (\ref{dftfree}) generates
a functional $\tilde{F}=F/(k_B T \liqr)$ that can be written
\begin{eqnarray}
\tilde{F}
&=&\int d\vec{r}~\left[
\frac{1}{2}n^2(\vec{r})
-\frac{w}
{6}
n^3(\vec{r}) + 
\frac{u}{12}
n^4(\vec{r}) \right]-\nonumber \\
& &\frac{1}{2} \int\int d\vec{r}~d\vec{r}_2~n(\vec{r})
C_2(|\vec{r}-\vec{r}_2|) n(\vec{r}_2).
\label{pfcfree}
\end{eqnarray}
The expansion coefficients $w$ 
and $u$ 
are treated as free parameters to
provide additional model flexibility,
and $n(\vec{r})$ is also typically allowed
to assume nonzero average values $n_0$.
Such an expansion is justified for kernels $C_2$
that produce low-amplitude
density profiles, which in general requires suppression of large wavenumber
two-body correlations.
The coefficient $w$ will be set to zero throughout this article,
as the $u n^4(\vec{r})/12$ term automatically gives rise to the
effective terms $u n_0 n^3(\vec{r})/3$ and $u n_0^2 n^2(\vec{r})/2~$
\cite{foot1}.
It will also be implied that $u=3$ throughout this study.

Three dynamic equations for $n(\vec{r})$ will be considered here.
The first is a purely diffusive Model B form,
\be
\frac{\partial n(\vec{r})}{\partial t} = \nabla^2\frac{\delta 
\tilde{F}}
{\delta n(\vec{r})}
\label{eq:pfcdyn1}
\ee
where $t$ is dimensionless time.
The second equation of motion introduces a faster inertial, quasi-phonon 
dynamic component in addition to diffusive dynamics \cite{mpfc}, 
\be
\frac{\partial^2 n(\vec{r})}{\partial t^2}+\beta\frac{\partial n(\vec{r})}
{\partial t}= \alpha^2 \nabla^2 \frac{\delta 
\tilde{F}}
{\delta n(\vec{r})}
\label{pfcinertia}
\ee
where $\alpha$ and $\beta$ are constants
related to sound speed and damping rate, respectively.
The final equation of motion, applicable to 
Eq.\ (\ref{dftfree}), is
\be
\frac{\partial \ln{(n(\vec{r})+1)}}{\partial t} = 
-\frac{\delta 
\tilde{F}}
{\delta n(\vec{r})}. 
\label{gupta}
\ee
With imposed density
conservation this equation provides an accelerated path to local energy minima
\cite{dasgupta92}.
All simulations in this study were performed in 3D using 
pseudo-spectral algorithms and periodic boundary conditions.
Those that employed Eqs.\ (\ref{eq:pfcdyn1}) or (\ref{pfcinertia}) 
used semi-implicit time stepping, while
those that employed Eq.\ (\ref{gupta}) used explicit time stepping.

\section{Order and Defects in PFC Models}
\label{sec:models}
In this section, the primary factors controlling defect stability
in PFC and CDFT models are first outlined through an analysis
of stacking faults applicable to both model types.
Three potential solutions to the problem of defect instability
are examined and shown to be sufficient for
stabilization of stacking faults in fcc crystals. These approaches
center on
multi-peaked correlation functions, few-peaked correlation
functions with broad effective envelopes, and entropy-driven formulations,
respectively.
Section \ref{sec:defects} expands on these concepts with
a more detailed examination of specific issues concerning 
defect structure-geometry, interactions, and dynamics.

\subsection{Large wavenumber or multi-peaked models}
\label{subsec:large}
The efficiency and tractability of PFC models relative to CDFT
are direct consequences of the central PFC approximation; truncation
of two-body correlations beyond the first few primary correlation 
peaks in Fourier space
(which permits the truncated expansion of the logarithm in 
Eq.\ (\ref{dftfree})).
All structural information is consolidated into the wavenumber range
around the first few primary peaks in the structure factor. This
still permits control of basic structural symmetries, but produces local
density peaks with broad, sine-wave like profiles rather than the sharply
peaked Gaussian profiles that emerge from CDFT functionals with
large wavenumber correlations.
Though this small wavenumber approximation reduces model
complexity and increases efficiency by orders of magnitude, it can
in some cases also reduce defect stability, as will be shown 
in the present subsection.
Methods by which defects can be stabilized while still retaining the
small wavenumber approximation
are discussed in subsection \ref{subsec:broad}.

\subsubsection{Many-peaked XPFC model}
\label{subsubsec:manyXPFC}
The effect of large wavenumber correlations on defect stability 
can be demonstrated through an examination of 
the PFC model formulation of Greenwood et al.\ 
\cite{nikpfcstruct10,nikpfcstruct11,xpfcbin11} (XPFC).
This approach allows one to stabilize many different crystal symmetries 
by constructing a correlation kernel 
with peaks, typically Gaussians,
located at the first few primary reflections of a given lattice structure.
The Fourier transformed XPFC kernel can be written
\be
\hat{C}_2(k)_i = 
-r + H
e^{-(k-k_i)^2/(2\alpha_i^2)}
e^{-\sigma^2 k_i^2/(2\rho_i \beta_i)}
\label{xpfckernel}
\ee
where $i$ denotes a family of lattice planes at wavenumber $k_i$,
and $\sigma$ is a temperature parameter.
The constants $\alpha_i$, $\rho_i$, and $\beta_i$ are the
Gaussian width (which sets the elastic constants), atomic density,
and number of planes, respectively,  
associated with the $i$th family of lattice planes.
We have introduced the additional constants $r$ and $H$ to permit
$\hat{C}_2(k)$ to assume negative values between peak maxima, which
improves numerical stability when used with Eq.\ (\ref{dftfree}).
The envelope of all selected Gaussians $i$ 
shifted by the constant $-r$ then
composes the final $\hat{C}_2(k)$.
For example, two such peaks 
with ratio $k_2/k_1=\sqrt{4/3}$
produce an fcc structure in which the first peak corresponds to
$\{111\}$ and the second to $\{200\}$, the two primary fcc reflections.

\begin{figure}[btp]
 \centering{
 \includegraphics*[width=0.2\textwidth,trim=550 0 0 0]{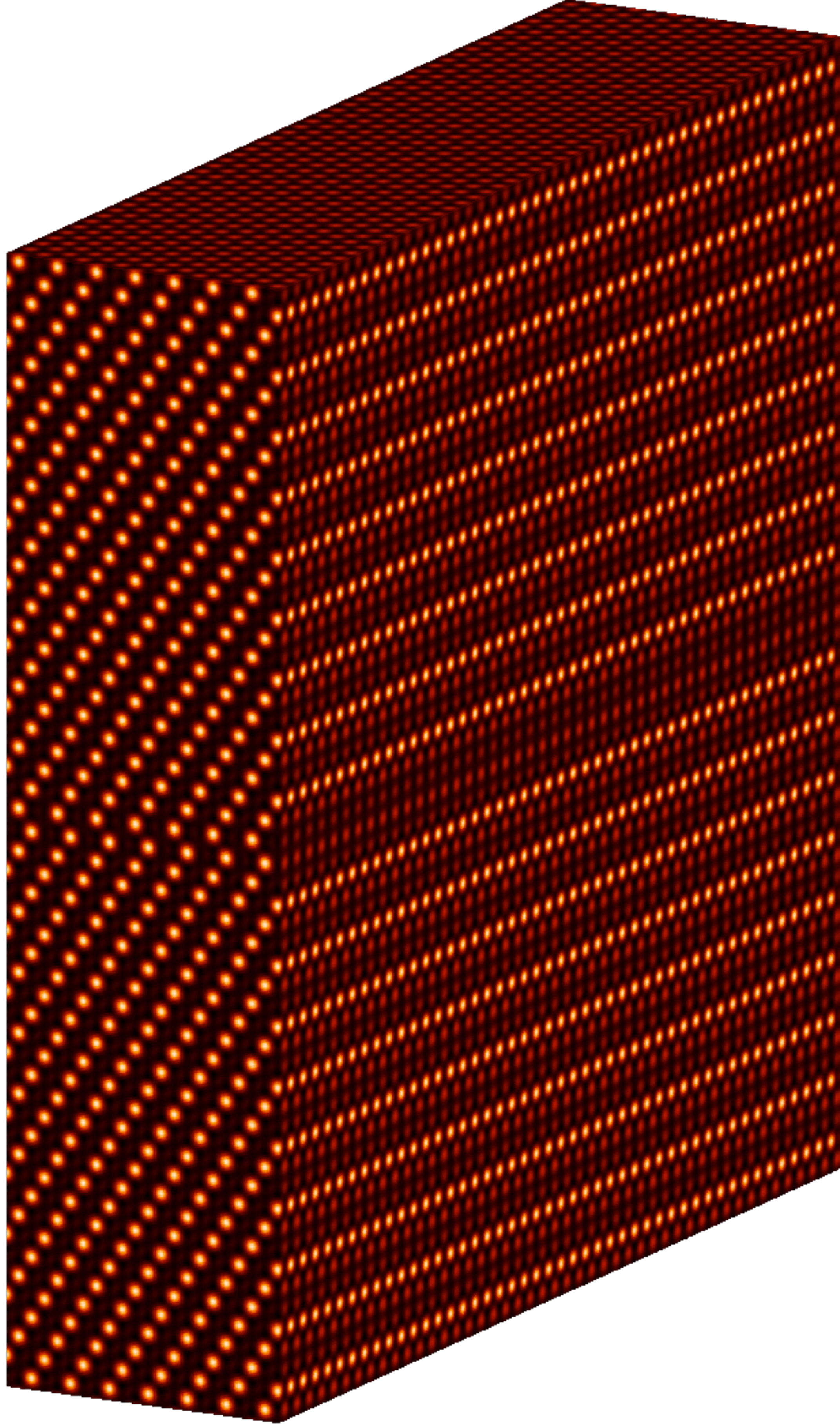}
   \put(-48,-22){(a)}
   \put(50,-22){(b)}
   \put(119,-22){(c)}
   \put(-26,-6){\begin{rotate}{7}\vector(4,-1){14}\end{rotate}}
   \put(-26,-6){\vector(2,1){14}}
   \put(-26,-6){\vector(0,1){14}}
   \put(-9,-12){\footnotesize $\vec{x}=[\bar{1}12]$}
   \put(-9,-1){\footnotesize $\vec{y}=[110]$}
   \put(-23,9){\footnotesize $\vec{z}=[1\bar{1}1]$}
~~~~~~~~~~~~~~~~~~~~~~
 \fbox{\includegraphics*[width=0.05\textwidth,trim=0 0 0 0]{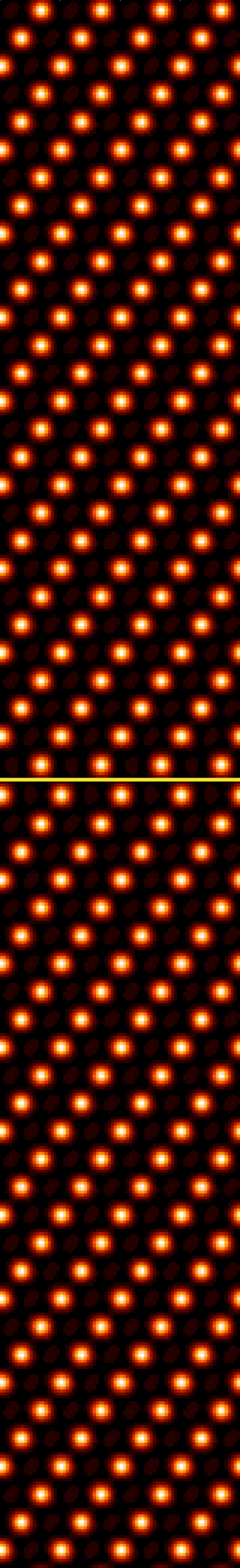}}
   \put(-16,-10){$\vec{x}$}
   \put(-35,81){$\vec{z}$}
   \put(-85,44){
 \fbox{\includegraphics*[width=0.08\textwidth,trim=123 908 40 900]{fig1b}}
   \put(-26,-10){$\vec{x}$}
   \put(-52,72){\footnotesize A}
   \put(-52,62){\footnotesize B}
   \put(-52,52){\footnotesize C}
   \put(-52,42){\footnotesize A}
   \put(-52,32){\footnotesize C}
   \put(-52,22){\footnotesize A}
   \put(-52,12){\footnotesize B}
   \put(-52,2){\footnotesize C}}
~~~
 \fbox{\includegraphics*[width=0.05\textwidth,trim=0 0 0 0]{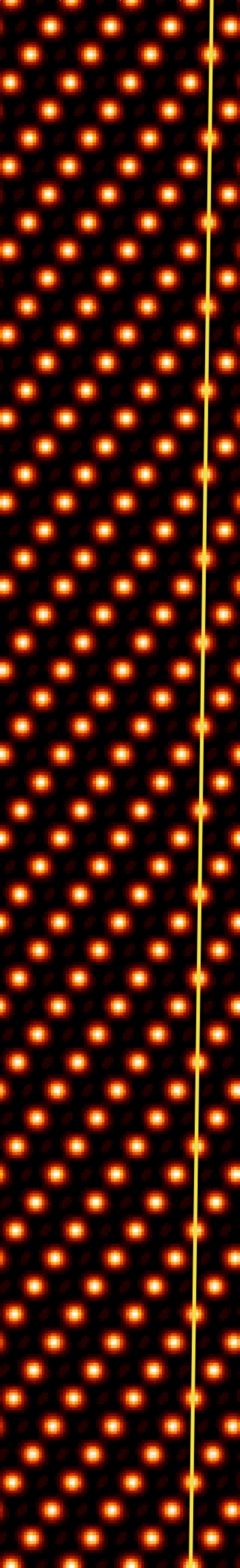}}
   \put(-16,-10){$\vec{x}$}
   \put(-35,81){$\vec{z}$}
 }
 \vspace{.05cm}
\caption[]
{\label{sfshear}
(Color online)
(a) 
An intrinsic $N_L=35$ fcc stacking fault in a periodic simulation cell;
(b) 
Close-up and full $xz$ views of the fault;
(c) 
Full $xz$ view of a sheared/unfaulted crystal in the same cell.
Yellow lines are guides to the eye.
These images were generated using Eqs.\ (\ref{pfcfree}) and
(\ref{spfckernel}) with 
parameter values $n_0=-0.48$, $r=-0.63$, and $B^x=1$.
}
\end{figure}

The issue of defect stability and its reliance on the correlation kernel
can be illustrated with an analysis of the fcc intrinsic stacking fault energy 
($\gamma_{\rm ISF}$ or SFE)
and stability as a function of the number of fcc reflections considered
in $\hat{C}_2(k)$. 
The full CDFT of Eq.\ (\ref{dftfree}) 
with $\hat{C}_2(k)$ given by Eq.\ (\ref{xpfckernel}) was employed,
as discussed below.
Faulted crystals were evaluated within a periodic simulation
cell with $(\vec{x},\vec{y},\vec{z})$ axes in 
$([\bar{1}12],[110],[1\bar{1}1])$ directions, and
a single intrinsic stacking fault was created 
by removing one close-packed layer and shifting the layers above down
by $a/\sqrt{3}$, where $a
=2\sqrt{3}\pi/k_1=1.8537$
is the equilibrium fcc lattice constant. 
The final number of close-packed layers $N_L$ must satisfy 
$N_L=3n-1$, where $n>1$ is an integer ($N_L=35$ was used in most cases here).
Initial states were relaxed using Eq.\ (\ref{gupta}).
In this geometry, the stacking fault may be stable, metastable, or
unstable, with instability unfolding through a shearing operation 
$\tau_{zx}$ or $\tau_{zy}$
that removes the stacking fault and creates a uniformly sheared perfect crystal
with total shear strain $\epsilon=(\sqrt{2} N_L)^{-1}$,
as shown in Fig.\ \ref{sfshear}.

\begin{figure}[btp]
 \setlength{\fboxrule}{.5pt}
 \centering{
 \includegraphics*[width=0.48\textwidth]{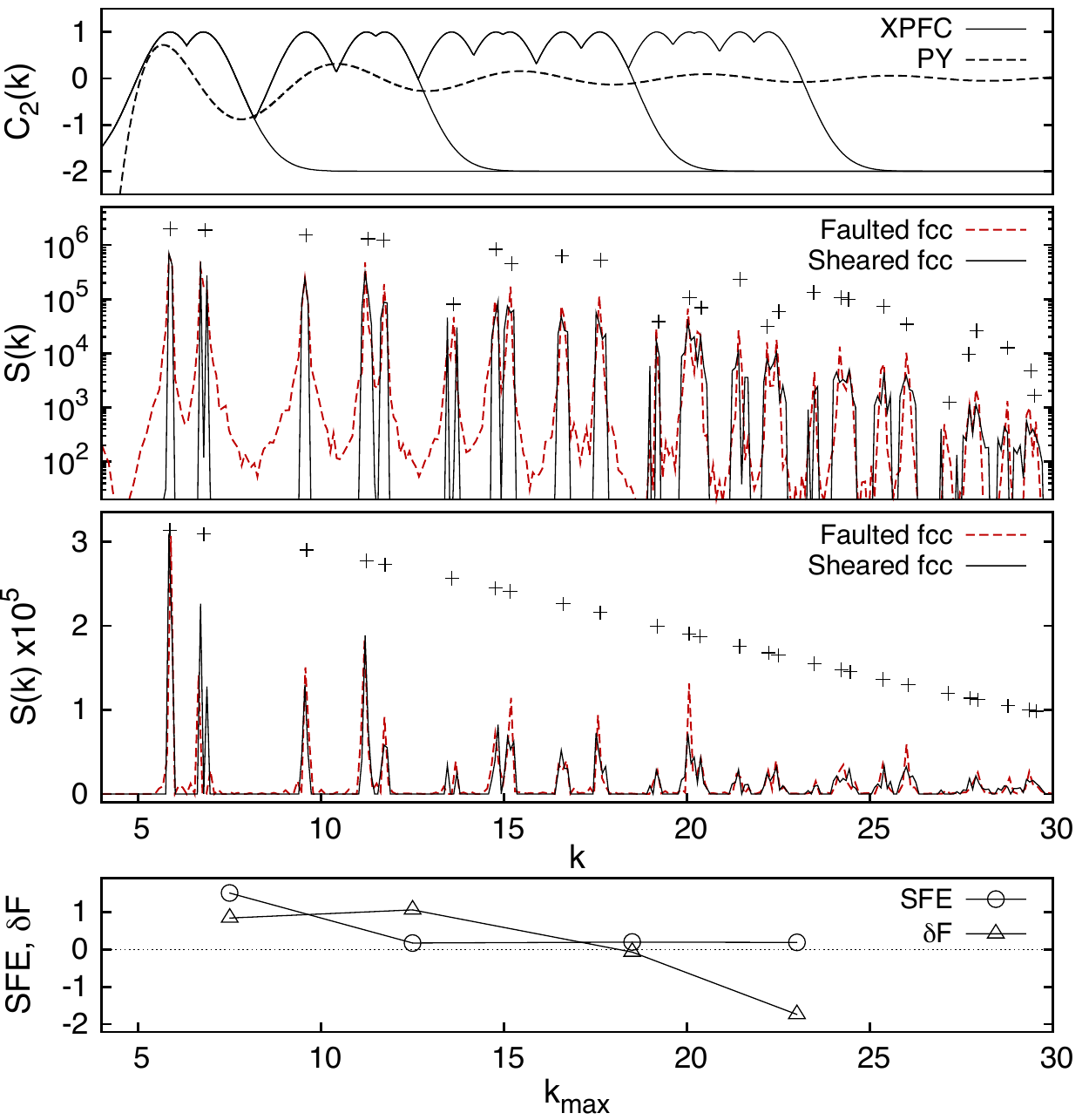}
   \put(-239.4,221){\begin{rotate}{90}$\hat{~}$\end{rotate}}
   \put(-239,47){\begin{rotate}{90}$\tilde{~}$\end{rotate}}
   \put(-33.5,40.6){\footnotesize$\tilde{~}$}
   \put(-225,252){
    \fbox{\includegraphics*[width=0.08\textwidth,trim=105 105 105 105]
     {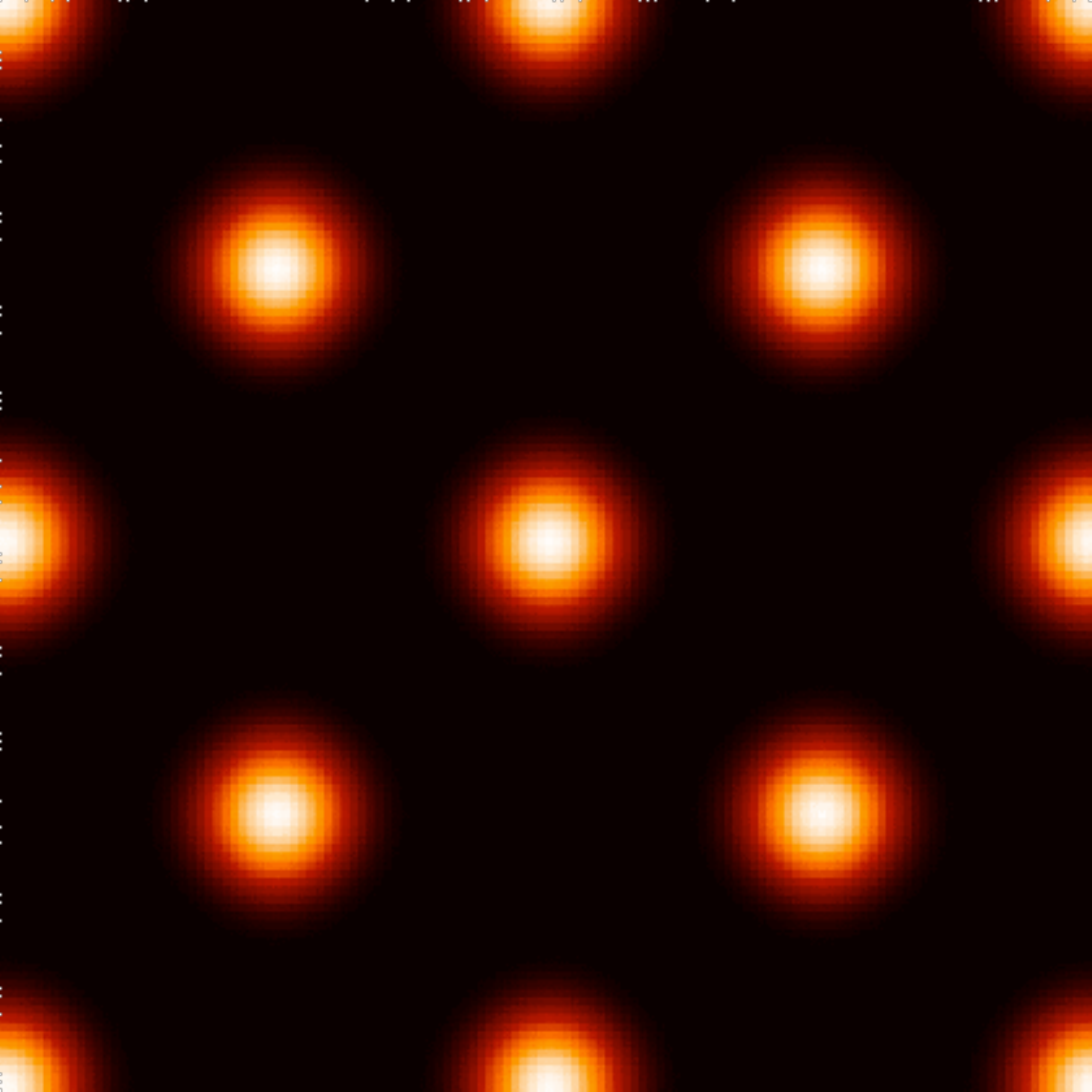}}}\put(-167.5,252){
    \fbox{\includegraphics*[width=0.08\textwidth,trim=105 105 105 105]
     {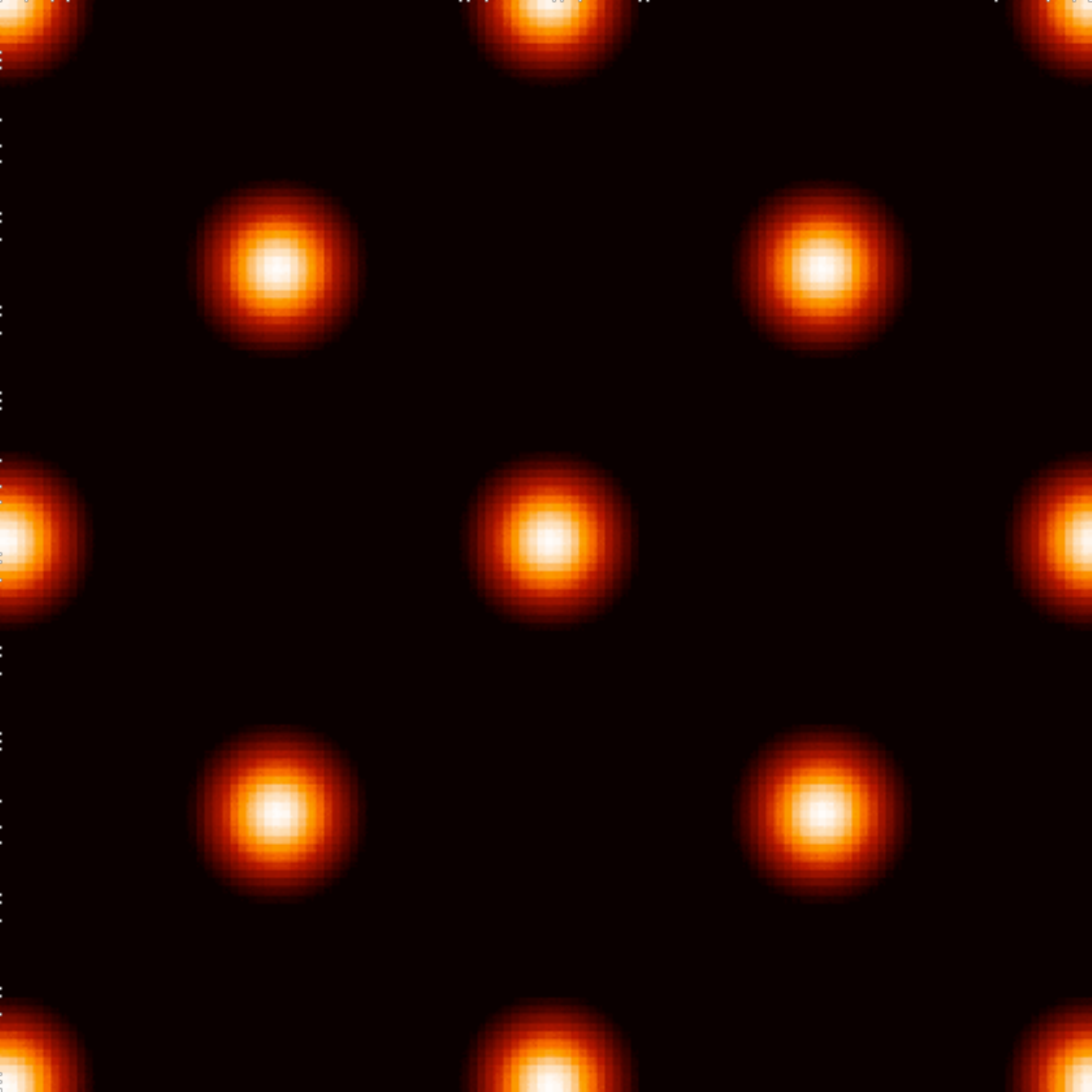}}}\put(-110,252){
    \fbox{\includegraphics*[width=0.08\textwidth,trim=105 105 105 105]
     {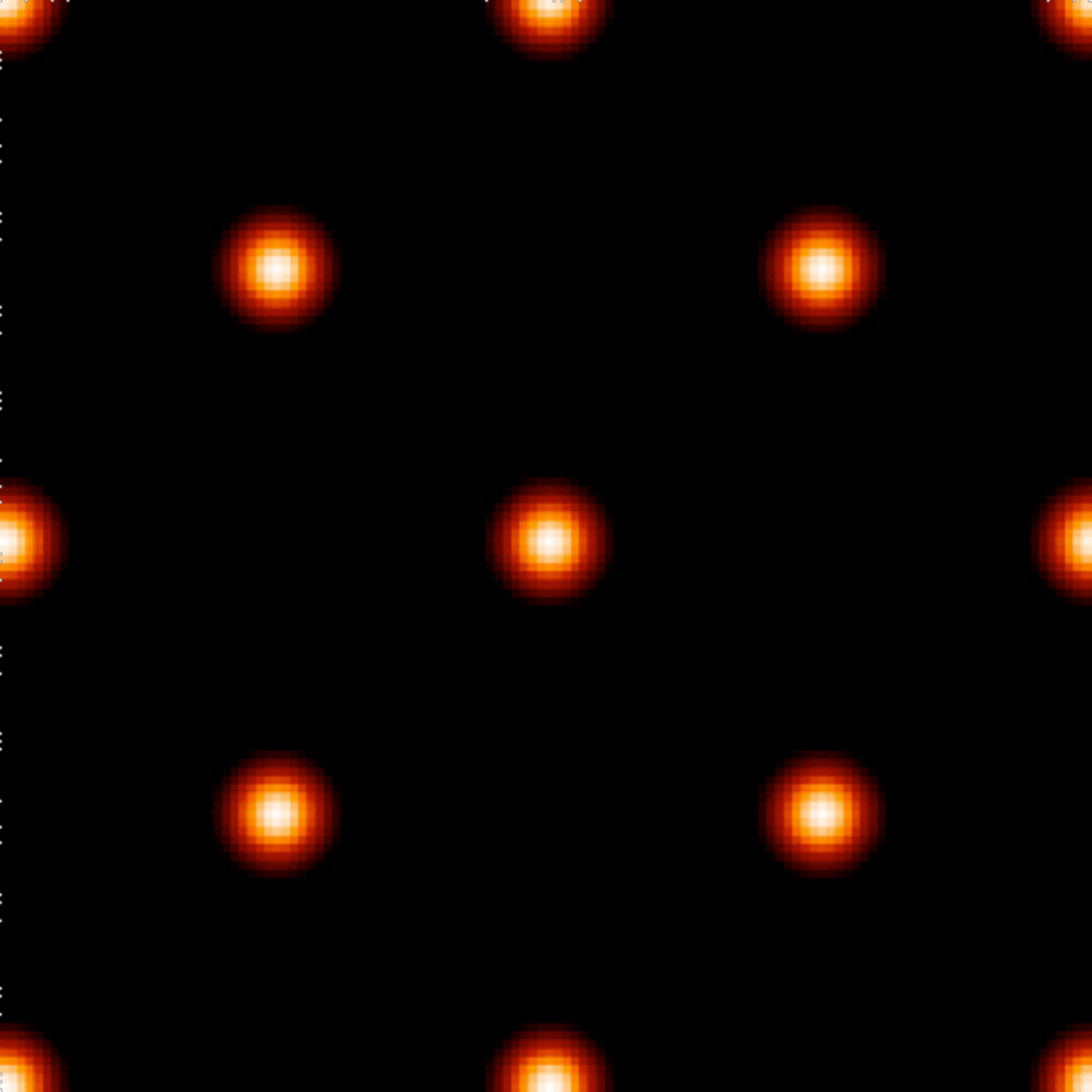}}}\put(-52.5,252){
    \fbox{\includegraphics*[width=0.08\textwidth,trim=105 105 105 105]
     {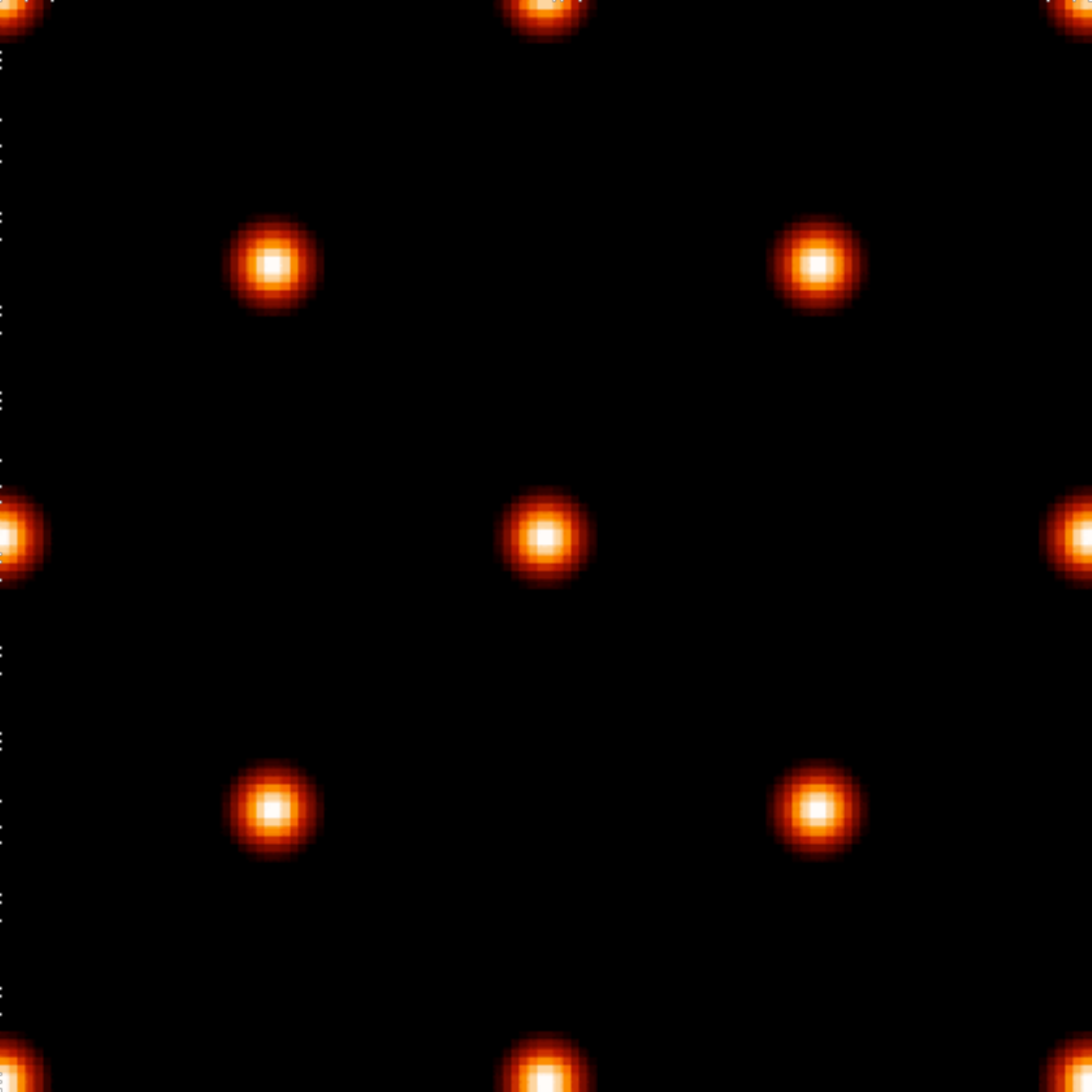}}}
 }
 \setlength{\fboxrule}{.5pt}
 \vspace{-.5cm}
\caption[]
{\label{xpfcpyc2}
(Color online)
Effect of large-$k$ modes on stacking fault stability.
From top to bottom:
Cross-sectional normalized $n(\vec{r})$ maps of the XPFC fcc crystal with 
2, 5, 10, and 15 fcc reflections from left to right;
$\hat{C}_2(k)$ of the PY hard-sphere model at $\liqr=0.9445$ and 
of the XPFC model with 2, 5, 10, and 15 fcc peaks;
Log plot of the structure factor 
$S(k)=\langle |\hat{n}(k)|^2 \rangle$
for faulted and sheared fcc crystals
in the 15 peak XPFC model;
Linear $S(k)$ plot for faulted and sheared fcc crystals
in the PY hard-sphere model;
XPFC $\gamma_{\rm ISF}$ and $N_L=35$ 
driving force for 
instability,
$\delta 
\tilde{F}
=
\tilde{F}
_{\rm SF}-
\tilde{F}
_{\rm Shear}$ vs.\ $k_{\rm max}$.
Symbols in the $S(k)$ plots are Bragg reflection maxima of
perfect fcc crystals in either model.
XPFC parameter values: $\liqr=1.0488$, $\alpha_i=1$, $\sigma=0$,
$H=3$, and $r=2$. $H$ and $r$ were varied from their default values,
$H=1$ and $r=0$, to ensure
numerical stability of the logarithmic term in Eq.\ (\ref{dftfree}).}
\end{figure}

Some results are displayed in Fig.\ \ref{xpfcpyc2} 
along with specific parameter values employed.
Stacking faults were found to be highly unstable in the two-peaked model
for virtually all meaningful values of $\alpha_i$ ($\lesssim 2$).
With five peaks, metastable faults were observed, while ten peaks
produced fully stable faults for which the energy of the faulted crystal
was very slightly lower than that of the sheared state
($\delta 
\tilde{F}
=
\tilde{F}
_{\rm SF}-
\tilde{F}
_{\rm Shear} \lesssim 0$). 
A 15 peak model 
was also found to be fully stable against shear, 
with a larger negative $\delta 
\tilde{F}
$. 
The relevant structure factors for the 15 peak model are shown in
Fig.\ \ref{xpfcpyc2}.
These results indicate that the stability of stacking faults to shear
increases as larger-$k$ correlations are considered.
The reason for this behavior is discussed later in this section.

Significantly, the computational efficiency of these models decreases 
rapidly with increasing $k_{\rm max}$. The amplitude of the density peaks
must be allowed to grow roughly as $k_{\rm max}^{e}$ if the
larger-$k$ modes are to contribute significantly to the free 
energy of the system. This necessitates
use of the full logarithmic one-body term in the 
free energy, without expansion, since the truncated expansion
of Eq.\ (\ref{pfcfree})
limits stable amplitudes to relatively small values, $\mathcal{O}(1)$.
The logarithm greatly reduces the maximum allowable time step, while
the higher-$k$ modes reduce the density peak width
(as shown in Fig.\ \ref{xpfcpyc2}, top) and therefore the maximum
allowable grid spacing $\Delta x$.

\subsubsection{Many-peaked CDFT model}
\label{subsubsec:manyCDFT}
\begin{figure}[btp]
 \setlength{\fboxrule}{.75pt}
 \centering{
~~~~~~~~~~~~~~~~~~~~~~
 \includegraphics*[width=0.34\textwidth]{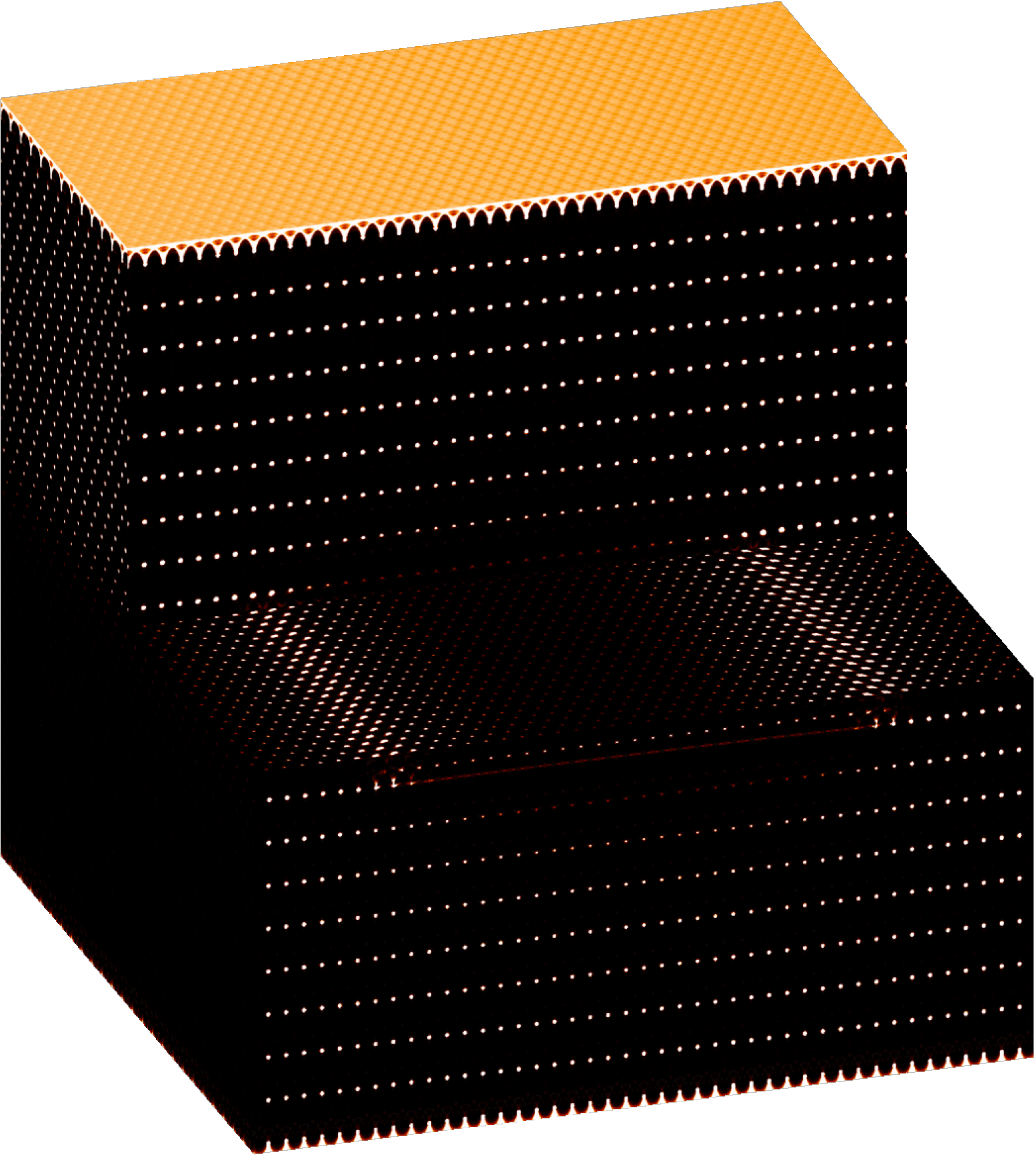} 
   \put(-242,58){\fbox{\includegraphics*[width=0.12\textwidth,trim=218 252 222 300]
    {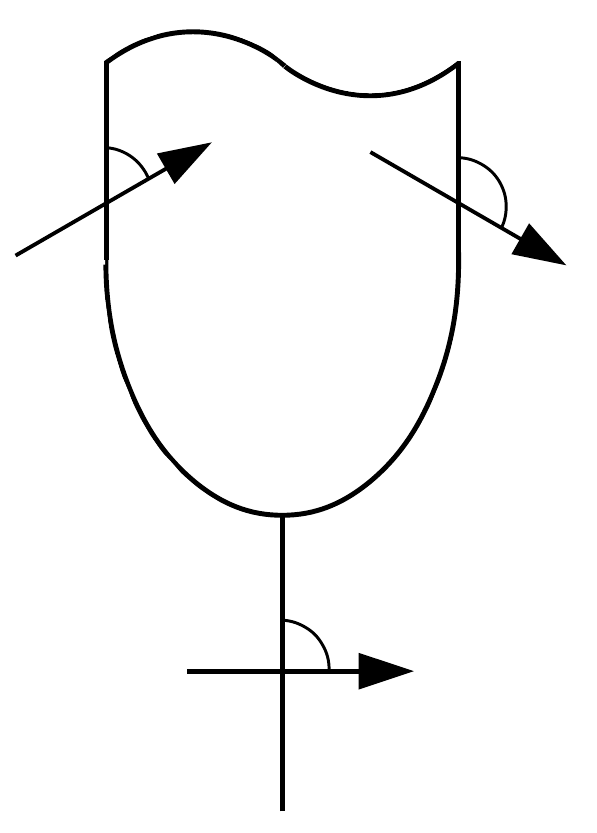}}
     \put(-39.5,46){\large \color{gray} SF}
     \put(-18,15){$\vec{b}$}
     \put(-30,23){$\theta$}
     \put(-10,48){$\vec{b}_1$}
     \put(-11,70){$\theta_1$}
     \put(-61,48){$\vec{b}_2$}
     \put(-61,70){$\theta_2$}}
   \put(-233,14){\vector(1,-1){10}}
   \put(-233,14){\vector(4,1){14}}
   \put(-233,14){\vector(0,1){14}}
   \put(-218,1){\footnotesize $\vec{x}=[\bar{1}12]$}
   \put(-216,16){\footnotesize $\vec{y}=[110]$}
   \put(-230,29){\footnotesize $\vec{z}=[1\bar{1}1]$}
   \put(-93,85){\footnotesize \color{white} \begin{rotate}{0} 
                  {Stacking} \end{rotate}}
   \put(-86,77){\footnotesize \color{white} \begin{rotate}{0} 
                  {Fault} \end{rotate}}
   \put(-103,135){\footnotesize \color{white} \begin{rotate}{0} {Shockley}\end{rotate}}
   \put(-101,127){\footnotesize \color{white} \begin{rotate}{0} {Partials}\end{rotate}}
   \put(-102.5,127){\color{white}\begin{rotate}{-8}\line(-1,-2){19}\end{rotate}}
   \put(-70,130){\color{white}\begin{rotate}{-5.5}\line(1,-1){26}\end{rotate}}
   \put(-137,72){\footnotesize \color{white} \begin{rotate}{0} $\vec{b}_2$\end{rotate}}
   \put(-126,75){\color{white}\vector(1,1){12}}
   \put(-23,88){\footnotesize \color{white} \begin{rotate}{0} $\vec{b}_1$\end{rotate}}
   \put(-47,94){\color{white}\vector(4,-1){20}}
 }
\caption[]
{\label{hssplit}
(Color online)
Split fcc edge dislocation in the 
$\liqr=0.9445$
PY hard-sphere CDFT
of Eq.\ (\ref{dftfree}).
A cutaway view of the density field $\rho(\vec{r})$ is shown,
with the color scale
truncated above $\rho(\vec{r})=1$, though 
$\rho(\vec{r}) \simeq 7800$ at the peak maxima.
Left: Diagrammatic representation of the dissociation of a perfect
dislocation $\vec{b}$ into Shockley partials $\vec{b}_1$ and $\vec{b}_2$.
}
\end{figure}
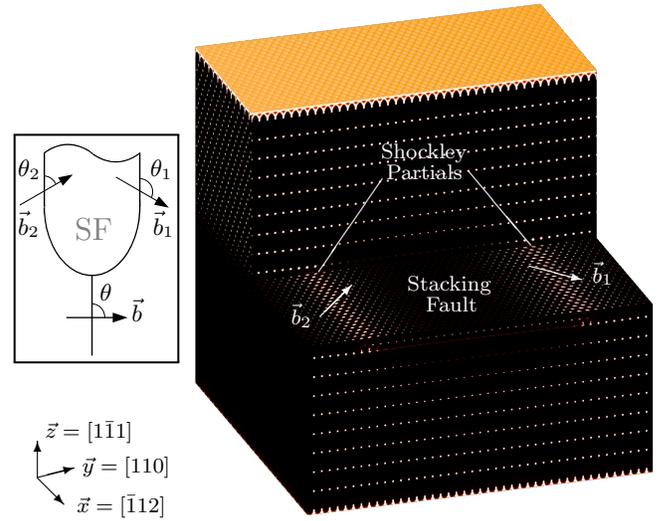

A similar analysis was attempted using the full RY
CDFT of Eq.\ (\ref{dftfree}),
with $\hat{C}_2(k)$ given by the Percus-Yevick (PY) expression
for hard spheres \cite{py58,pyhs63}
and dynamics given by Eq.\ (\ref{gupta}).
Our findings indicate that hcp crystals in this model actually tend
to have a slightly lower free energy than fcc crystals, 
a fact that apparently has not been appreciated before. 
This leads to a negative
$\gamma_{\rm ISF}$ and therefore stacking faults that are inherently stable to
the sheared state. 
Nonetheless, a perfect edge dislocation in a metastable fcc crystal 
was found to split into two well-separated Shockley partial 
dislocations connected by a stacking fault, as expected for low 
$\gamma_{\rm ISF}$
materials (Fig.\ \ref{hssplit}). The reaction is
\be
\vec{b}=\frac{a}{2}[110] \rightarrow 
\frac{a}{6}[121] + \gamma_{\rm ISF} +
\frac{a}{6}[21\bar{1}].
\label{reaction}
\ee

Though both $\gamma_{\rm ISF}$ and $\delta 
\tilde{F}
$ in this system are 
always negative or very small and positive, 
making instability to shear virtually impossible, 
we did find that
$\delta 
\tilde{F}
$ increases as $k_{\rm max}$ is reduced.
This was determined by modifying the PY function as 
$\hat{C}_2(k)=\hat{C}_2^{\rm PY}(k) e^{-(k-k_{\rm max})/2}$ for 
$k \ge k_{\rm max}$, while
holding $\Delta x$ constant. 
The observed trend in $\delta 
\tilde{F}
$ indicates that the underlying link
between stability and $k_{\rm max}$ found in the XPFC model is also present here.

\subsubsection{Analysis and generalization}
\label{subsubsec:general}
This link can be
understood by examining the spectra of the two relevant states, 
the faulted crystal and the sheared/unfaulted crystal 
(see Fig.\ \ref{xpfcpyc2}).
An 
infinite
stacking fault can be expressed mathematically as a 1D step
function in the 3D displacement field $u(\vec{r})$ of the crystal, 
with the step direction normal to the fault.
The structure factor of a faulted crystal therefore contains the
same $1/k^2$ line-shapes characteristic of scattering from a surface.
In this case each lattice reflection $k^{(i)}$, except the set of $\{111\}$
planes parallel to the stacking fault, will possess its own set of
$1/(k_z-k_z^{(i)})^2$ modes, and these modes will only extend in the reciprocal
space direction perpendicular to the plane of the fault, $k_z$.   
The spherically averaged structure factor $S(k)$ thus exhibits a sequence
of broadened fcc reflections, each with effective $1/(k-k^{(i)})^2$ line-shapes
of roughly the same width \cite{foot2}.

The structure factor of a sheared perfect crystal also exhibits a
form of peak broadening, but in this case the degree of broadening is
proportional to $k$. 
Depending on the orientation of shear strain $\epsilon$, 
or deformation in general,
the spacing $d$ between certain lattice planes within any given family
changes to some value $d'$. 
A simple analysis shows that the resulting shift in $k$-space under shear is
\be
\delta k = 2\pi \left( \frac{1}{d'}-\frac{1}{d} \right) =
\frac{1-\cos{\omega}}{\cos{\omega}} k
\label{shearshift}
\ee
where $\omega=\arctan{\epsilon}$.
Thus the primary low-$k$ reflections remain sharp and/or slightly split,
while the degree of shift or effective broadening increases in proportion
to $k$ due to this growing decoherence effect across high-index, low-$d$ 
planes. 

The driving force for stacking fault instability
is primarily a function of 
$\delta 
\tilde{F}
=
\tilde{F}
_{\rm SF}-
\tilde{F}
_{\rm Shear}$.
When a low-mode kernel is used,
$
\tilde{F}
_{\rm Shear}$ will tend to be smaller than $
\tilde{F}
_{\rm SF}$ 
because
the slightly shifted low-$k$ modes of a sheared crystal induce a
small energy cost relative to that of the long-range line-shape tails
of a faulted crystal,
provided that the shear strain is not too large.
As $k_{\rm max}$ increases, this situation eventually reverses. 
The 
high order
reflections from
the sheared lattice become broader or less coherent than those of the 
faulted crystal such that if enough high-$k$ modes are considered, 
stacking fault stability can eventually be attained when 
$
\tilde{F}
_{\rm Shear} > 
\tilde{F}
_{\rm SF}$.
This effect can be seen in the data of Fig.\ \ref{xpfcpyc2}.
We emphasize that the high-$k$ modes do not necessarily decrease
$\gamma_{\rm ISF}$ or $
\tilde{F}
_{\rm SF}$, 
but that the increase in $
\tilde{F}
_{\rm Shear}$ alone
leads to a dramatic increase stacking fault stability.
In the geometry of our simulations, the critical wavenumber 
at which $
\tilde{F}
_{\rm SF}=
\tilde{F}
_{\rm Shear}$ increases
with $N_L$ since $\epsilon=(\sqrt{2}N_L)^{-1}$, but for fixed $N_L$
stability increases with $k_{\rm max}$.
The exact wavenumber at which stability is attained 
($k_{\rm max} \simeq 18$ in this case) also depends on model
parameters, but the basic trends should generally hold. 

\begin{figure}[btp]
 \centering{
 \includegraphics*[width=0.48\textwidth]{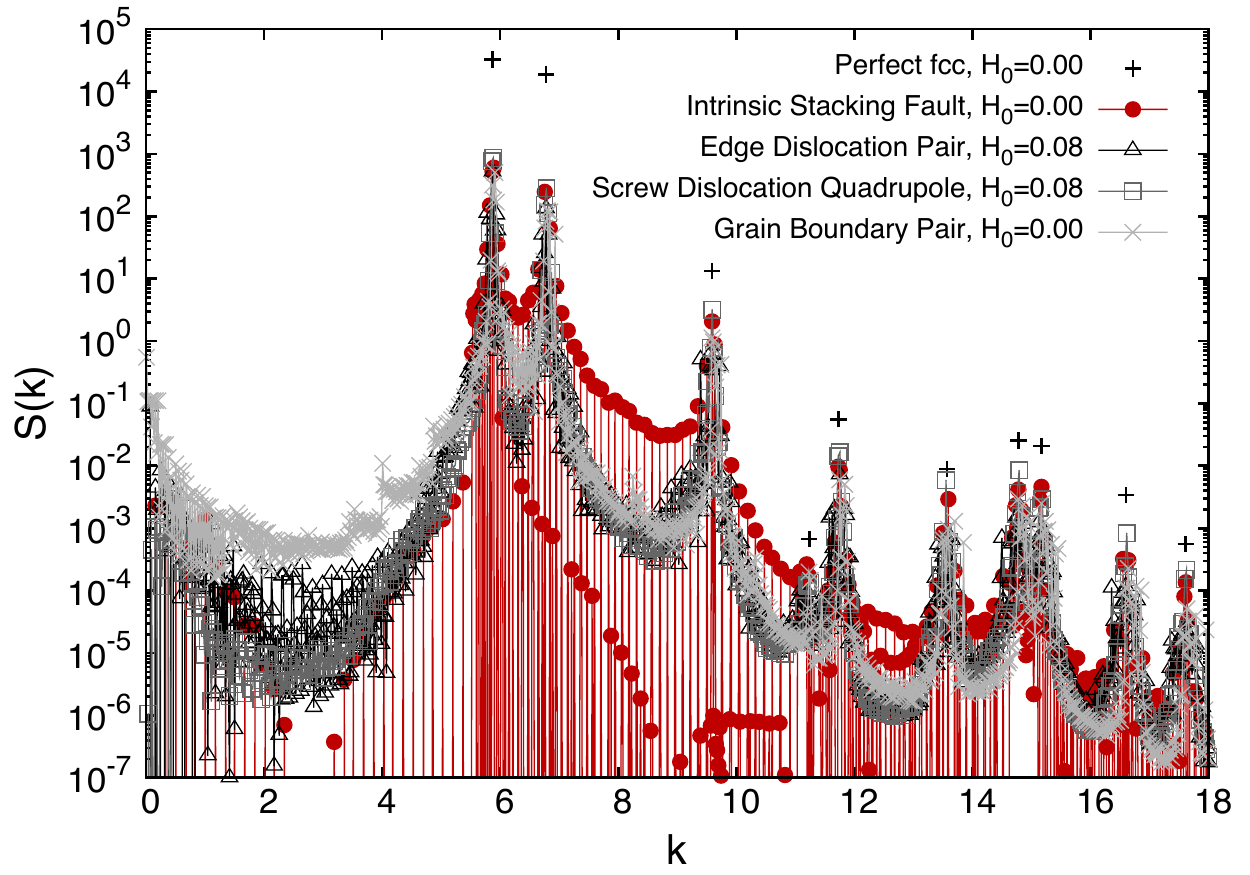} 
 }
 \vspace{-.25cm}
\caption[]
{\label{defectspectra}
(Color online)
Structure factors of fcc crystals with various types of defects.
Note that the most compact structure, the planar stacking fault,
produces the overall broadest line-shapes.
These spectra were generated using Eqs.\ (\ref{pfcfree}) and
(\ref{pfcmod}) with 
parameter values $n_0=-0.48$, $r=-0.63$, $B^x=1$, $\alpha_0=1/2$,
and $k_0=6.2653$.}
\end{figure}

Similar arguments can be constructed for other types of defects
by reconsidering for each case the two generic parameters controlling stability,
driving force and energy barrier.
The driving force for instability is controlled by
both the energy of the defected state and
the energies of the states resulting from defect removal.
In the case just examined, this force was quantified by $\delta 
\tilde{F}
$,
the energy difference between the faulted state and the competing sheared
state. The magnitude of this force is therefore deeply tied to 
the precise form of $\hat{C}_2(k)$ and the energy cost that it imposes
on defect line-shapes, 
a few examples of which are shown in Fig.\ \ref{defectspectra}.
These spectra were obtained using the modulated version of the original 
PFC model presented in Section \ref{subsec:spfc}, though the basic
features do not depend on the model used.
The four crystals shown have roughly equivalent defect densities
and their structure factors exhibit similar line-shapes,
with the stacking fault producing the greatest degree of broadening.
It is apparent that in general, 
broad correlation kernels will reduce the energies of all of
these defect structures and therefore the driving force for their instability,
as demonstrated 
explicitly
in the next subsection.
Alternatively, stacking faults were stabilized in the previous example by 
instead increasing the energy of the post-defect-removal sheared state 
through high-$k$ correlation effects. This simply illustrates that the 
driving force can be controlled by varying the energy of 
the initial state, the final state, or both through the form of
$\hat{C}_2(k)$.

The second parameter relevant to stability, 
the energy barrier for defect removal or topological protection,
is controlled by the nature of
the process by which a given defect can be removed.
The shearing operation that removes a stacking fault requires only small
local translations of the density peaks. No peaks are created or removed,
thus no long-range mass transport is involved, and the resulting energy
barrier is small.
Stability in such a case therefore requires that the driving force be minimized,
since the energy barrier is in general more difficult to manipulate.
Removal of dislocations and grain boundaries on the other 
hand requires more elaborate transformations that often involve climb, 
with redistribution of atoms by long-range diffusive mass transport 
(adding or removing density peaks in the case of a PFC model).
These defects thus have greater inherent stability than stacking faults
due to both their somewhat narrower line-shapes and 
their generally larger energy barriers for removal \cite{foot3}.

The results presented in this section demonstrate that 
certain defect structures can be stabilized in PFC models
by building structural information back into the model
with additional correlation peaks. 
The resulting multi-peaked models require the full logarithmic free
energy functional and therefore suffer from marked
inefficiencies and greater complexity relative to existing PFC models
studied in the literature.
However, the sources of defect instability uncovered here will prove
useful in analyzing and constructing
few-peaked PFC models that both retain efficiency and stabilize relevant
defects.

\subsection{Small wavenumber or low-mode models}
\label{subsec:broad}
The inefficiencies associated with large wavenumber models can be avoided
by maintaining the small wavenumber PFC approximation and giving closer
treatment to the details of the one or few correlation peaks employed.   
We have successfully stabilized stacking faults using four variations
of PFC with the small wavenumber approximation, though in general this
requires some degree of parameter tuning and/or slight modification of
$\hat{C}_2(k)$.

Perhaps the simplest way to improve defect stability in this approximation
is to use a broad 
$\hat{C}_2(k)$ 
in the region
of the first few primary lattice reflections.
This in general lowers the energy associated with defect line-shapes
while having little to no effect on the energy of, for example, uniformly
sheared states or undeformed crystals.
A sufficiently broad envelope can thus stabilize
the $1/k^2$ stacking fault modes relative to the first few
sheared state $\delta k$ shifts. 
The main drawback is that the stability of the equilibrium crystal
structure tends to decrease as 
$\hat{C}_2(k)$ 
is broadened.
For example, a two-peaked XPFC fcc model can prefer hcp, bcc, rod, or
lamellar structures over fcc for sufficiently large $\alpha_i$.
Thus defect stability is not given freely, it comes inversely bound
to crystal stability. 
This trade-off must be managed by ensuring that the primary
reflections of the desired crystal structure continue to be preferred
over those of competing symmetries that may exist within a broad
$\hat{C}_2(k)$ 
envelope.
Specific parameter values related to
the width of 
$\hat{C}_2(k)$ 
(elastic moduli) therefore
become restricted, but 
a workable
balance between defect stability and crystal stability is still often
attainable in the small wavenumber PFC approximation,
as demonstrated in the following subsections. 

\subsubsection{Original PFC model}
\label{subsubsec:SPFC}
\begin{figure}[btp]
 \centering{
 \includegraphics*[width=0.48\textwidth]{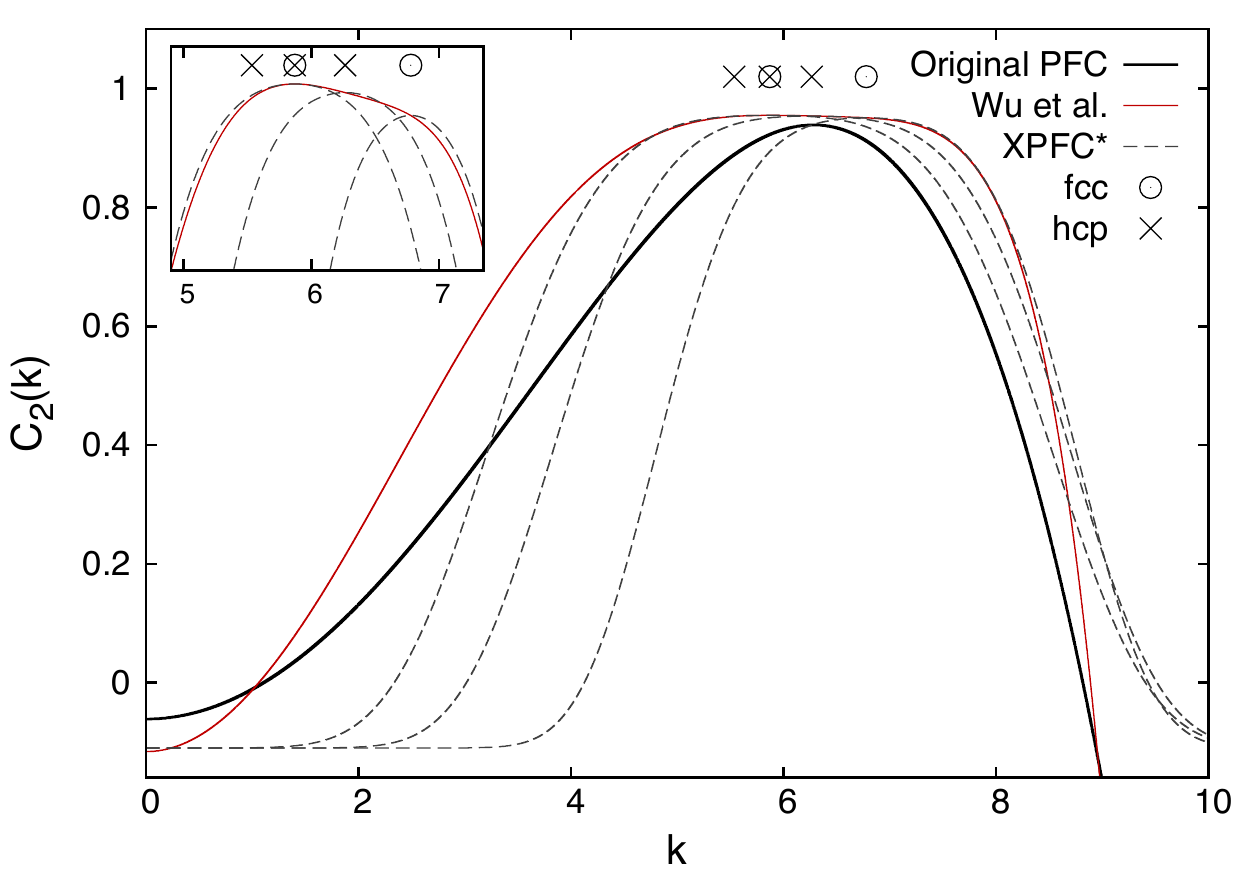} 
   \put(-239,84.5){\begin{rotate}{90}$\hat{~}$\end{rotate}}
 }
 \vspace{-.25cm}
\caption[]
{\label{c2model}
(Color online)
Sample PFC 
kernels
that produce stable stacking faults,
with relevant fcc and hcp reflections also indicated.
Inset: Close-up view of the peak maxima.
Original PFC parameters: $n_0=-0.48$, $r=-0.63$ and $B^x=1$.
${\rm XPFC}^*$ parameters, Eq.\ (\ref{xpfckernel2}): 
$n_0=-3/10$, $r=-9/40$, $w_2=1/50$, $w_4=49/50$, $\alpha_i \simeq 1$,
$\sigma=0$, and $H \simeq 1.0625$.
Wu et al.\ parameters: $n_0=-3/10$, $r=-9/40$, $R_1=1/20$, and $B^x=1$.
Note that for display purposes
all kernels have been shifted by $-3n_0^2$ to maintain
consistency with Eq.\ (\ref{dftfree}), and that
$k$ has been rescaled in the ${\rm XPFC}^*$ and Wu et al.\ kernels 
to match the equilibrium fcc reflections of the original PFC kernel.}
\end{figure}

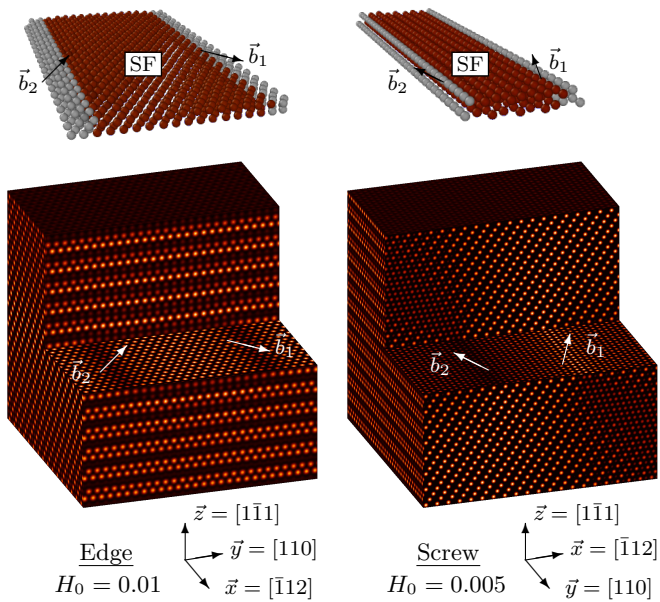
\begin{figure}[btp]
 \setlength{\fboxrule}{0.5pt}%
 \setlength{\fboxsep}{1.5pt}%
 \centering{
 \includegraphics*[width=0.23\textwidth]{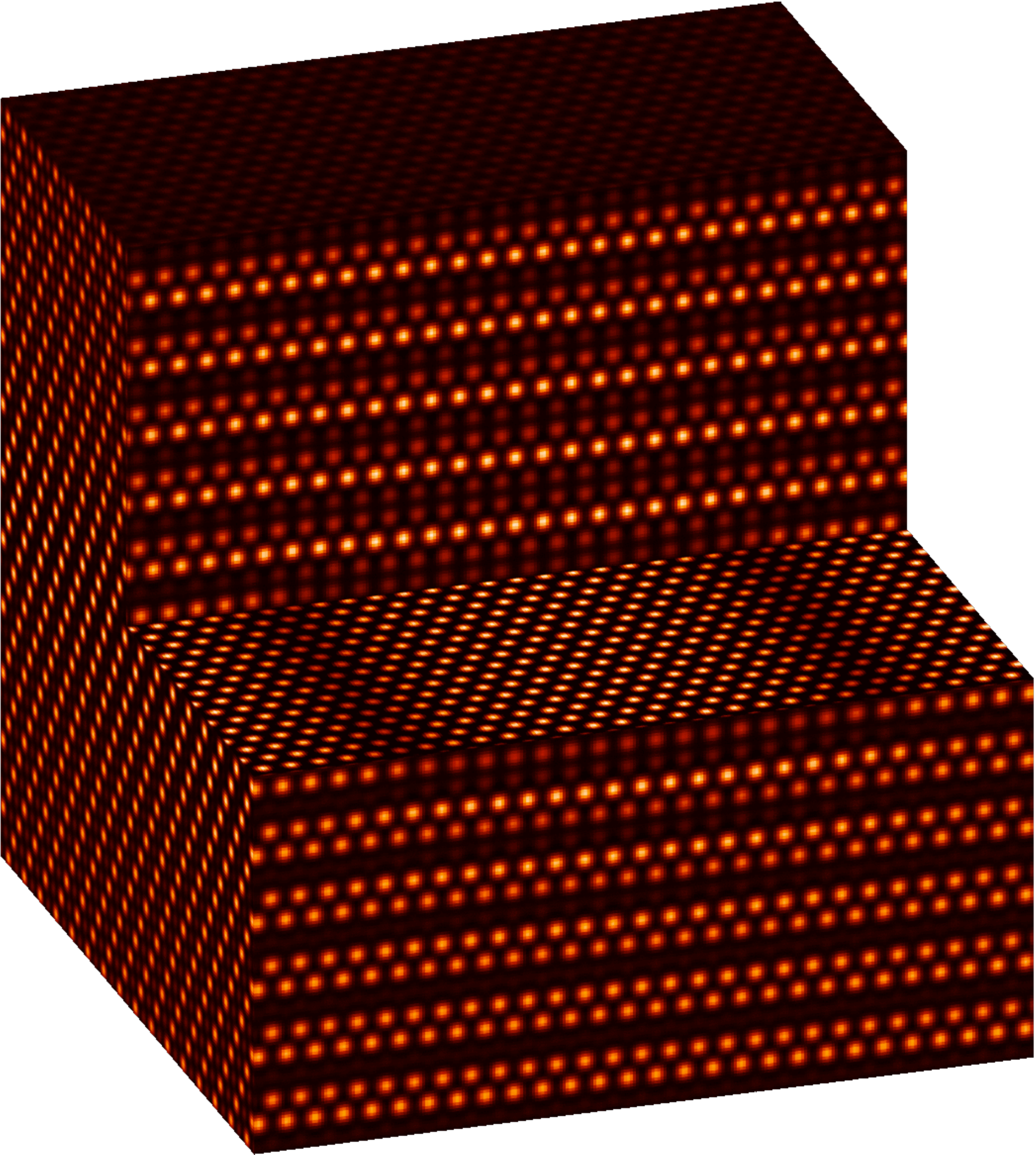}
   \put(-120,125){\includegraphics*[width=0.23\textwidth,trim=260 300 280 200]{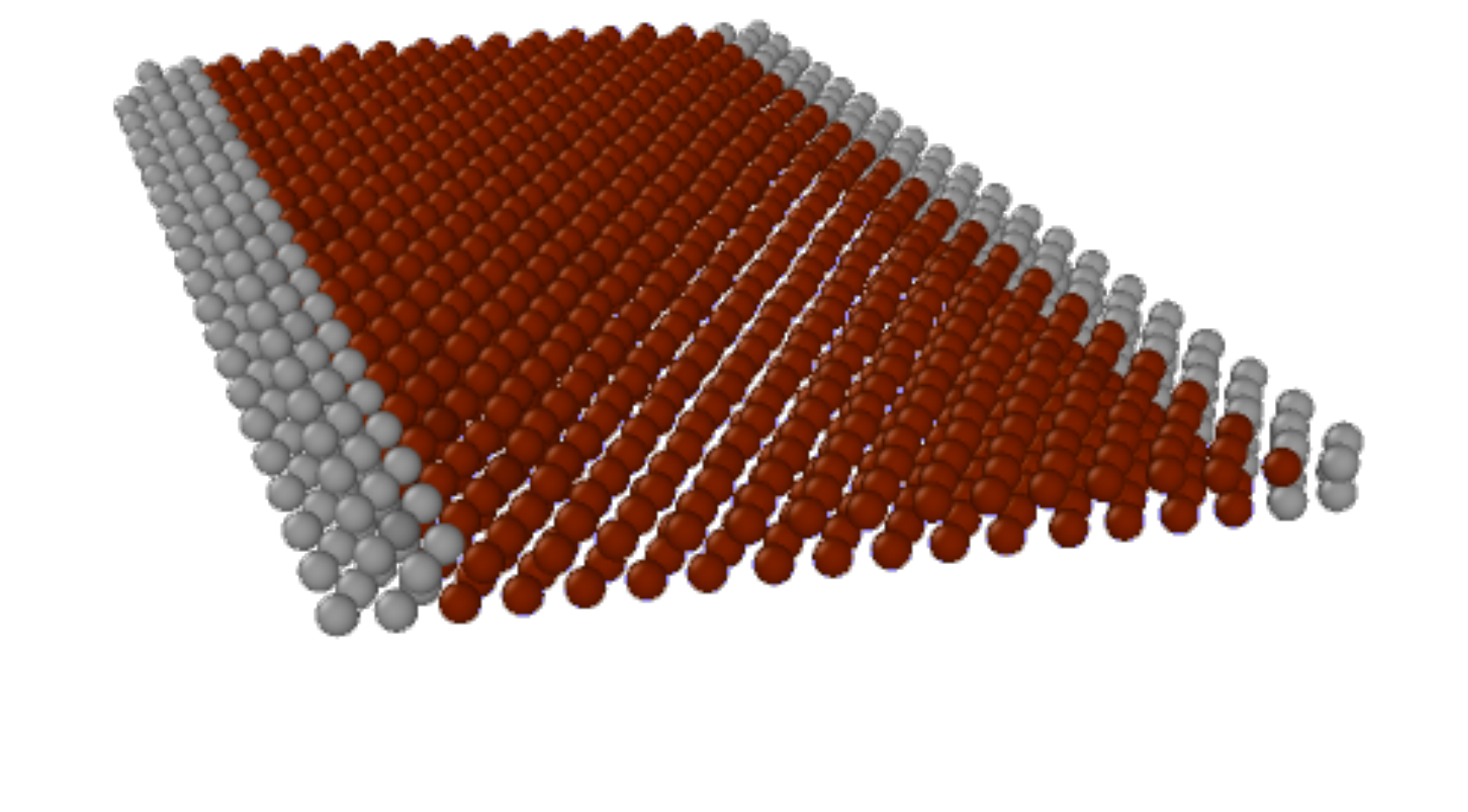}}
   \put(-73,165){\fcolorbox{black}{white}{\footnotesize SF}}
   \put(-113,158){\footnotesize $\vec{b}_2$}
   \put(-104,161){\vector(1,1){10}}
   \put(-26,168){\footnotesize $\vec{b}_1$}
   \put(-44,173){\vector(4,-1){16}}
   \put(-50,-20){\begin{rotate}{-5}\vector(1,-1){10}\end{rotate}}
   \put(-50,-20){\begin{rotate}{-5}\vector(4,1){14}\end{rotate}}
   \put(-50,-20){\vector(0,1){14}}
   \put(-35,-33){\footnotesize $\vec{x}=[\bar{1}12]$}
   \put(-33,-18){\footnotesize $\vec{y}=[110]$}
   \put(-47,-5){\footnotesize $\vec{z}=[1\bar{1}1]$}
   \put(-90,-20){\ul{Edge}}
   \put(-99.5,-32){$H_0=0.01$}
   \put(-93,48){\footnotesize \color{white} \begin{rotate}{0} $\vec{b}_2$\end{rotate}}
   \put(-82,51){\color{white}\vector(1,1){10}}
   \put(-16,58){\footnotesize \color{white} \begin{rotate}{0} $\vec{b}_1$\end{rotate}}
   \put(-34,63){\color{white}\vector(4,-1){16}}
~~~
 \includegraphics*[width=0.23\textwidth]{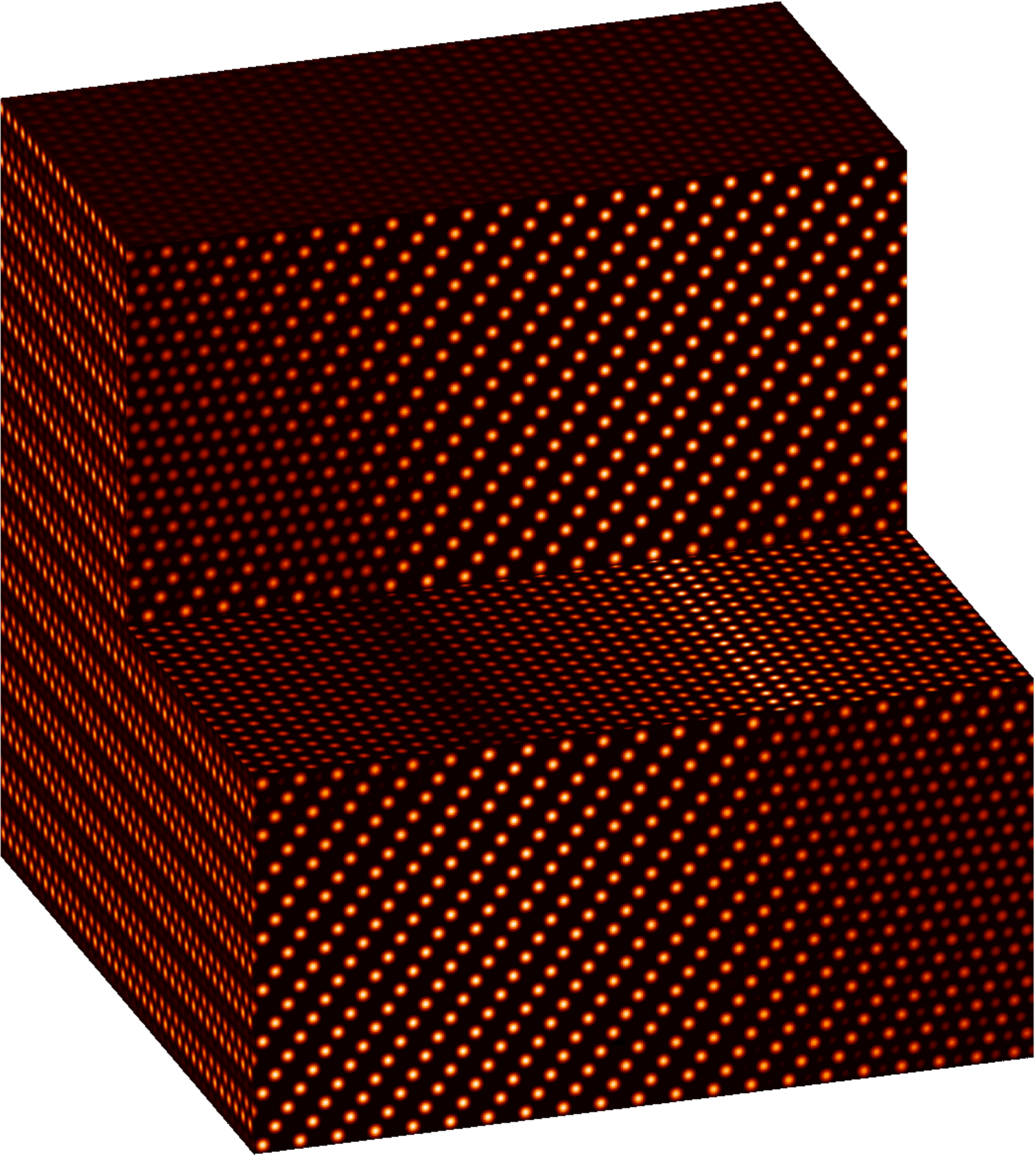} 
   \put(-120,125){\includegraphics*[width=0.23\textwidth,trim=300 300 300 200]{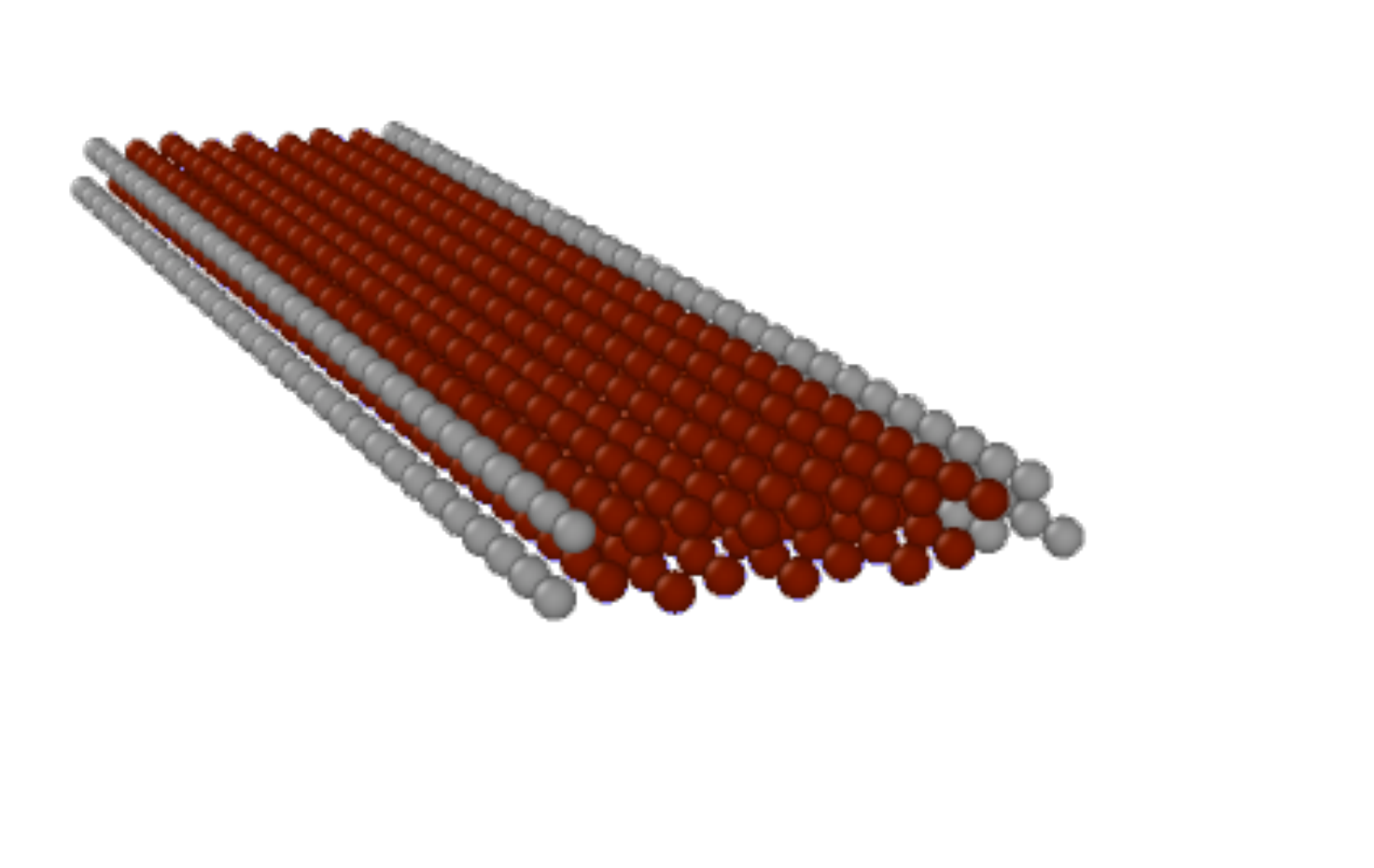}}
   \put(-78,165){\fcolorbox{black}{white}{\footnotesize SF}}
   \put(-102,158){\footnotesize $\vec{b}_2$}
   \put(-81,161){\vector(-2,1){11}}
   \put(-41,166){\footnotesize $\vec{b}_1$}
   \put(-44,162){\begin{rotate}{20}\vector(0,1){10}\end{rotate}}
   \put(-50,-20){\begin{rotate}{-5}\vector(1,-1){10}\end{rotate}}
   \put(-50,-20){\begin{rotate}{-5}\vector(4,1){14}\end{rotate}}
   \put(-50,-20){\vector(0,1){14}}
   \put(-35,-33){\footnotesize $\vec{y}=[110]$}
   \put(-33,-18){\footnotesize $\vec{x}=[\bar{1}12]$}
   \put(-47,-5){\footnotesize $\vec{z}=[1\bar{1}1]$}
   \put(-91,-20){\ul{Screw}}
   \put(-102.5,-32){$H_0=0.005$}
   \put(-86,51){\footnotesize \color{white} \begin{rotate}{0} $\vec{b}_2$\end{rotate}}
   \put(-64,52){\color{white}\vector(-2,1){13}}
   \put(-27,58){\footnotesize \color{white} \begin{rotate}{0} $\vec{b}_1$\end{rotate}}
   \put(-36,54){\begin{rotate}{-14}\color{white}\vector(0,1){12}\end{rotate}}
 }
 \vspace{-.05cm}
\caption[]
{\label{edgescrew}
(Color online)
Dissociated $\vec{b}=a/2[110]$ edge and screw dislocations in the 
(modulated) original PFC model 
of Eqs.\ (\ref{pfcfree}) and (\ref{pfcmod}).
The lower images show cutaway views of $n(\vec{r})$, illustrating
the mixed Shockley partials connected by a single intrinsic stacking
fault.
The upper images show atomic representations of the peak positions in
$n(\vec{r})$: red atoms have local hcp coordination
(stacking fault), gray atoms have local noncrystalline coordination
(defect cores), and atoms with fcc coordination are not shown.
Parameter values $n_0=-0.48$, $r=-0.63$, $B^x=1$, $\alpha_0=1/2$,
and $k_0=6.2653$ were used.
}
\end{figure}

We begin with the original PFC model of Elder et al.\ \cite{pfc02,pfc04},
for which the correlation kernel and the primary fcc and hcp
reflections are shown in Fig.\ \ref{c2model}.
This PFC or Brazovskii \cite{braz75} functional
specifies a kernel
\be
\hat{C}_2(k) = -r  + 1 - B^x (1-\tilde{k}^2)^2,
\label{spfckernel}
\ee
where $B^x$ is a constant proportional
to the solid-phase elastic moduli and $\tilde{k}=k/(2\pi)$.
This 
$\hat{C}_2(k)$ 
produces equilibrium fcc structures within a certain
parameter range \cite{pfc3dphases10,pfcnuclgranasy10}, but the energy of the hcp
crystal is necessarily always very close to that of fcc, and is sometimes 
lower.
This means that the fcc $\gamma_{\rm ISF}$ 
is always vanishingly small in the original PFC
model, which naturally stabilizes stacking faults relative to shear.
Perfect dislocations, as with the CDFT hard-sphere model, split into
two well-separated Shockley partials and a stacking fault, as shown in
Fig.\ \ref{edgescrew}. 
Note that, for display purposes, 
the defects shown in Fig.\ \ref{edgescrew} were generated using
a modified version of Eq.\ (\ref{spfckernel}), discussed in
Section \ref{subsec:spfc}.
The small value of $\gamma_{\rm ISF}$ produced by Eq.\ (\ref{spfckernel})
leads to an equilibrium splitting distance 
$\sim 32a$ for $\vec{b}=a/2[110]$ edge dislocations,
which is significantly larger than values $\lesssim 10a$
observed in typical single component fcc metals. 

The broad 
$\hat{C}_2(k)$ 
function of the original PFC model therefore naturally 
stabilizes defects,
but is relatively limited in terms of flexibility in controlling 
crystal symmetry and elastic properties. One is restricted to the deep
quench region of fcc stability, where small changes to the elasticity parameter
$B^x$ tend to destabilize fcc, and the inherent $\gamma_{\rm ISF}$ is 
extremely small.
It is possible to tune $\gamma_{\rm ISF}$ within a small range by
varying the envelope width $B^x$, but a more effective approach is to
explicitly enhance or suppress specific hcp and fcc modes, 
as demonstrated in Section \ref{sec:defects}.

\subsubsection{Low-mode XPFC model}
\label{subsubsec:fewXPFC}
XPFC models with only a few low-$k$ peaks can also be tuned to support 
less protected defect structures,
though some modifications seem to be necessary in the case of fcc
stacking faults.
As $\alpha_i$ is increased above $\sim 1$ or $2$, Gaussian kernel peaks
generally become too broad to maintain a consistent crystal structure
before they are broad enough near their
maxima to sufficiently support the $1/k^2$ stacking fault modes.
We have obtained better results in this regard using a modified Gaussian 
shape function,
\be
\hat{C}_2(k)_i = -r + H
e^{-\frac{w_2 (k-k_i)^2 + w_4 (k-k_i)^4}{2\alpha_i^2}}
e^{-\frac{\sigma^2 k_i^2}{2\rho_i \beta_i}}
\label{xpfckernel2}
\ee
where the coefficients $w_2+w_4=1$ set the relative weight of
$k^2$ versus $k^4$ shape functions.
This formulation still permits control of the phase diagram in the original 
spirit of the model as well as the magnitude and anisotropy of elastic 
constants, 
though the range of parameter values 
sufficient for
stability remains restricted.
Values in the vicinity of $w_2=1/50$, $w_4=49/50$, $\alpha_i=1$,
$\sigma=0$, 
$H=1$, $r=-9/40$, and $n_0=-3/10$
have been found sufficient to stabilize stacking faults in a three-peaked 
fcc model with two fcc reflections at $k_2=\sqrt{4/3}k_1$
and one commensurate hcp reflection at $k_3=\sqrt{41/36}k_1$ 
(see Fig.\ \ref{c2model}).
Similar results have been obtained using only the two fcc reflections
with their maxima connected by any nearly linear bridge function.

These findings highlight the conflict
between crystal stability and defect stability in small wavenumber models.
Very narrow crystal-stabilizing XPFC kernels destabilize defects of
all but the most protected types, somewhat broader kernels still tend
to destabilize susceptible planar fault structures, while very broad fully
defect-stabilizing XPFC kernels ultimately
resemble the simpler kernels of the original PFC model.
These inherent trade-offs suggest a limitation to the range
of atomic-level crystal and defect structures that low-mode, two-body
PFC models can capture, 
though the ultimate limit apparently lies at some level of complexity beyond
that of the primary defect structures in fcc crystals.

\subsubsection{Wu et al.\ fcc PFC model}
\label{subsubsec:wukarma}
The two-mode fcc model introduced by Wu et al.\ \cite{wukarmafcc} also 
produces stable stacking faults within certain parameter ranges.
To facilitate tuning of $\gamma_{\rm ISF}$ and stability, 
we find it helpful to retain a bulk modulus coefficient $B^x$ analogous
to that of the original PFC kernel,
\be
\hat{C}_2(k) = -r + 1 
- B^x (1-\tilde{k}^2)^2 \left[ 
(Q_1^2-\tilde{k}^2)^2 + \frac{R_1}{B^x} \right]
\label{wukarmakernel}
\ee
where $Q_1=\sqrt{4/3}$ is the wavenumber of the second mode
and $R_1$ sets its height. 
We find that quenches of intermediate depth tend to produce fcc crystals
with moderate $\gamma_{\rm ISF}$, and that $\gamma_{\rm ISF}$ decreases 
with increasing $|n_0|$.
This is because the energy of hcp approaches and may become lower than
that of fcc as the quench becomes deeper. 
As an example, the parameter set 
$r=-9/40$, $R_1=1/20$, $B^x=1$, and
$n_0=-1/4$
produces metastable stacking faults for $N_L=35$
and perfect dislocations that do not split into partials.
As 
$n_0$
is lowered, $N_L=35$ stacking faults eventually become fully stable,
with 
$n_0=-3/10$
producing a very low 
$\gamma_{\rm ISF}$ and widely split partial
dislocations (see Fig.\ \ref{c2model}).
$\gamma_{\rm ISF}$ also varies rapidly with $B^x$, 
again permitting only a small degree of elastic moduli tuning.

The source of defect stability in this PFC model is apparent from 
examination of the kernel around the primary fcc modes. 
Elasticity parameters $B^x \lesssim 1$ produce broad profiles with very
shallow wells between to the two modes, while $B^x \gtrsim 1$ produces
narrowing profiles with increasingly deeper wells. 
Using the parameter values above at 
$n_0=-3/10$
and $B^x=2$, 
the stacking fault line-shapes are already unfavorable enough to fully
destabilize $N_L=35$ faulted crystals.
This two-mode model comes with similar restrictions in terms of
acceptable parameter ranges to those of the original and XPFC models,
but does provide greater control of fcc stability than the original
PFC model, allowing use of shallower quenches closer to the linear elastic
regime.

\subsubsection{Vacancy PFC model}
\label{subsubsec:VPFC}
The vacancy PFC or VPFC model typically employs 
the same 
$\hat{C}_2(k)$ 
as the original 
PFC model, but adds a strong local nonlinear cutoff term that destroys the
inherent one-mode nature of inhomogeneous density states and produces
relatively autonomous individual density peaks separated by regions of
zero density \cite{vpfc09,vpfcglass11}.
This can lead to stabilization of individual lattice vacancies and in
general produces a Brownian hard-sphere-like system in which
fcc and hcp structures have lowest free energy, with a slight preference
toward fcc. When vacancies are stabilized in this way, packing effects become
more important and the natural similarities between fcc and hcp in this
regard should lead to low $\gamma_{\rm ISF}$ systems. 
We find that a low $\gamma_{\rm ISF}$, comparable to that of the original PFC kernel,
is indeed obtained with the VPFC description,
though in this case we attribute the low value primarily to entropic or 
packing effects. 
In this sense, the VPFC model is likely the
most defect-friendly of existing PFC models, but one must reinterpret
the time scales accessible to this description since the explicit vacancy
diffusion mechanism has been partially reintroduced.
Nonetheless, stacking faults are quite stable in this formulation
and perfect dislocations split properly into partials
joined by a stacking fault.

\section{Survey of fcc Defect Properties in a Small Wavenumber PFC Model}
\label{sec:defects}
The findings presented in Section \ref{sec:models} suggest that many
types of crystalline defects observed in real materials, even those 
with relatively low inherent stability, can be stabilized within
existing PFC models. The general requirement for stability under
the small wavenumber approximation involves a sufficiently broad correlation
kernel maximum, which 
leads to restrictions on the parameter ranges that provide 
both defect and crystal
stability.
Nonetheless, these restrictions appear to be manageable in the case of
close-packed crystals,
and closer examinations of fcc systems reveal that
a wide variety of realistic defect properties
naturally emerge within these parameter ranges. 
Some of these properties are discussed
in the present section.

Emphasis is placed on fcc crystals and tuning of the stacking fault energy
to generate either
undissociated perfect dislocations (high $\gamma_{\rm ISF}$) 
or dissociated Shockley partial
dislocations bound by stacking faults (low $\gamma_{\rm ISF}$).
Various fundamental properties of both edge and screw dislocations 
in these configurations are examined.

\subsection{Modulation of the original PFC model}
\label{subsec:spfc}
As discussed in Section \ref{subsubsec:SPFC},
the dissociation width of a perfect dislocation in the original PFC model 
is larger than that of most single component fcc metals,
due to the small value of $\gamma_{\rm ISF}$ relative to the fcc
elastic moduli.
This behavior can be tuned by various means.
Since $\gamma_{\rm ISF}$ is strongly linked to the difference in free 
energy between
commensurate fcc and hcp states, one method to control $\gamma_{\rm ISF}$ is to 
selectively promote or suppress the primary hcp reflections that do not
coincide with those of the fcc crystal. For example,
the commensurate hcp reflection at $
k_0
=2\pi\sqrt{41/12}/a$ can be suppressed
by subtracting a small 
XPFC-type
Gaussian centered at this wavenumber 
from the 
$\hat{C}_2(k)$ 
of Eq.\ (\ref{spfckernel}),
such that the two fcc reflections are not affected,
\be
\hat{C}_2(k) = -r + 1 - B^x (1-\tilde{k}^2)^2 -
H_0 e^{-(k-k_0)^2/(2\alpha_0^2)}.
\label{pfcmod}
\ee
The constants $H_0$ and $\alpha_0$ are analogous to those of 
Eq.\ (\ref{xpfckernel}).

\begin{figure}[btp]
 \centering{
 \includegraphics*[width=0.48\textwidth]{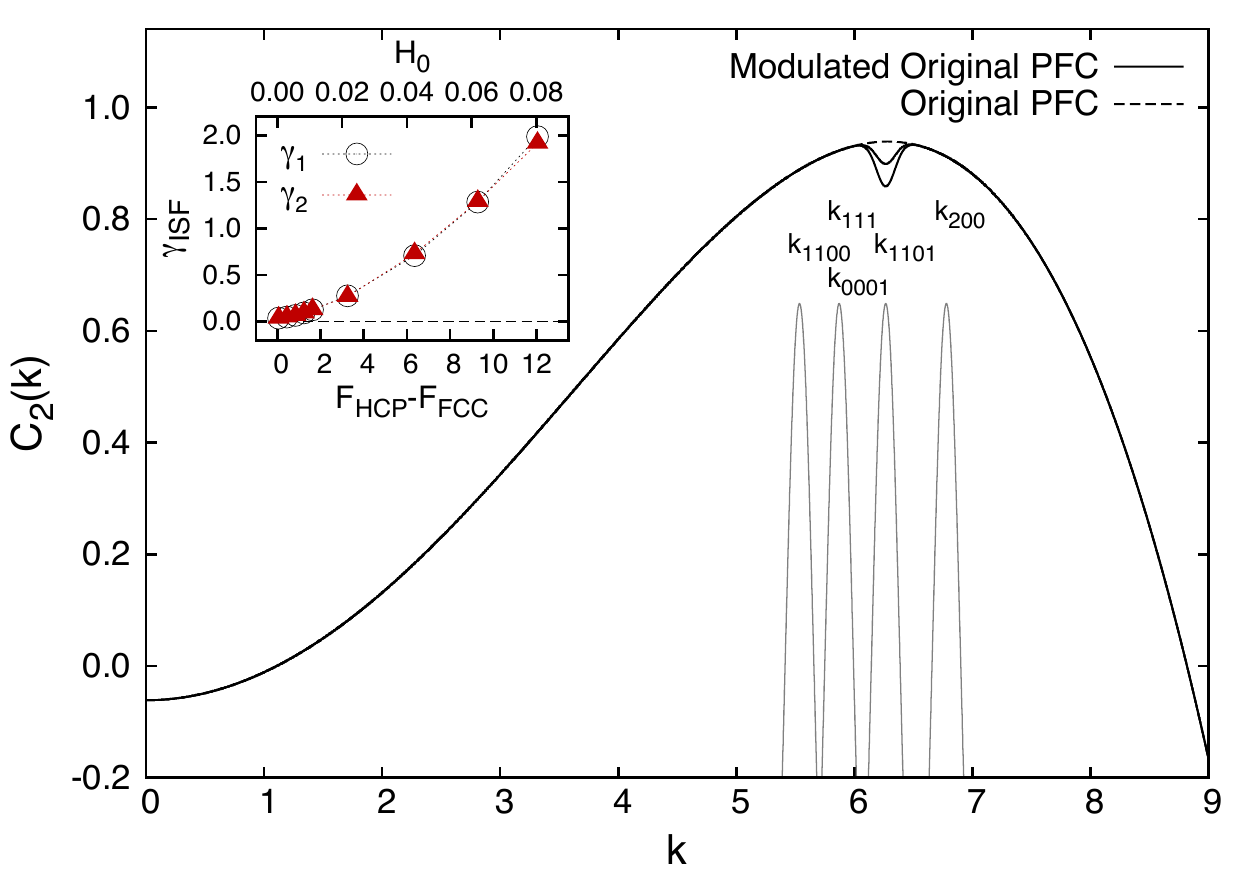} 
   \put(-239,84.5){\begin{rotate}{90}$\hat{~}$\end{rotate}}
   \put(-178.5,91.5){$\tilde{~}$}
   \put(-162.5,91.5){$\tilde{~}$}
 }
 \vspace{-.25cm}
\caption[]
{\label{c2gamma}
(Color online)
Modulated original PFC kernel of 
Eq.\ (\ref{pfcmod}) where
$n_0=-0.48$, $r=-0.63$, $B^x=1$, $\alpha_0=1/2$, and $k_0=6.2653$. 
$H_0=0$, $0.04$, and $0.08$ are shown.
Inset: 
$\gamma_{\rm ISF}$ obtained with the given modulated kernel vs.\
$
\tilde{F}
_{\rm hcp}-
\tilde{F}
_{\rm fcc}$ and $H_0$.
$\gamma_1$ corresponds to $N_L=53$ and $\gamma_2$ to $N_L=35$.
As in Fig.\ \ref{c2model},
all kernels have been shifted by $-3n_0^2$.
}
\end{figure}

The original PFC kernel modified in this way is shown in 
Fig.\ \ref{c2gamma}.
The result is greater stabilization of fcc due to the increase in 
$
\tilde{F}
_{\rm hcp}$,
and a corresponding increase in $\gamma_{\rm ISF}$ as the 
height of the subtracted Gaussian $H_0$ becomes larger.
This 
$\hat{C}_2(k)$ 
is similar to that of Wu et al.\ with variable $B^x$,
though some subtle differences are apparent.
Only $\gamma_{\rm ISF}$ of the equilibrium fcc state changes as $H_0$ is varied,
producing an essentially fixed reference system with tunable $\gamma_{\rm ISF}$.
The standard phase diagram
is of course modified for nonzero $H_0$,
proportionally expanding the region of fcc stability.
The dependence of $\gamma_{\rm ISF}$ on $H_0$ is also shown in 
Fig.\ \ref{c2gamma}. $N_L=35$ faults are stable or metastable 
for $H_0 \lesssim 0.05$, and
very little system size dependence is observed in $\gamma_{\rm ISF}$ for
$N_L \gtrsim 35$. 

\subsection{Dissociation width vs.\ $\gamma_{\rm ISF}$}
\label{subsec:deq}
This method of controlling $\gamma_{\rm ISF}$ can be used to examine
defect properties that depend on stacking fault energy, and to compare
functional dependencies with those predicted by continuum elastic theories.
For example, the elastic strain energy of a fcc crystal
containing a perfect dislocation with $b=a/\sqrt{2}$ 
is lowered by dissociation into two
Shockley partials with $b_p=a/\sqrt{6}$ and a stacking fault, as described by
Eq.\ (\ref{reaction}).
This is confirmed by the Frank criterion, $b^2=a^2/2 > 2 b_p^2=a^2/3$.
Continuum elastic theories provide predictions for
the equilibrium separation $d_{\rm eq}$ between the two resultant
Shockley partials. 
The long-range elastic energies of the partials
produce a repulsive force per length of dislocation line proportional to $1/d$, 
while the energy cost of the fault increases with $d$,
producing an attractive force per length proportional to $\sfe$. 
The separation that balances these two forces
for an elastically isotropic material is given by
\cite{hirthlothe}
\be
d_{\rm eq}=\frac{2-\nu}{1-\nu}\left( 1-\frac{2\nu \cos{2\beta}}{2-\nu}
\right)\frac{\mu b_p^2}{8\pi \gamma_{{\rm ISF}}}
\label{deq}
\ee
where $\nu$ is the Poisson ratio, $\beta$ is the angle between the
Burgers vector of the perfect dislocation and its line direction,
and $\mu$ is the isotropic shear modulus of the crystal.
Similar though more complicated results can be derived using anisotropic
elasticity or various approximations thereof \cite{hirthlothe}.

\begin{figure}[btp]
 \centering{
 \includegraphics*[width=0.48\textwidth]{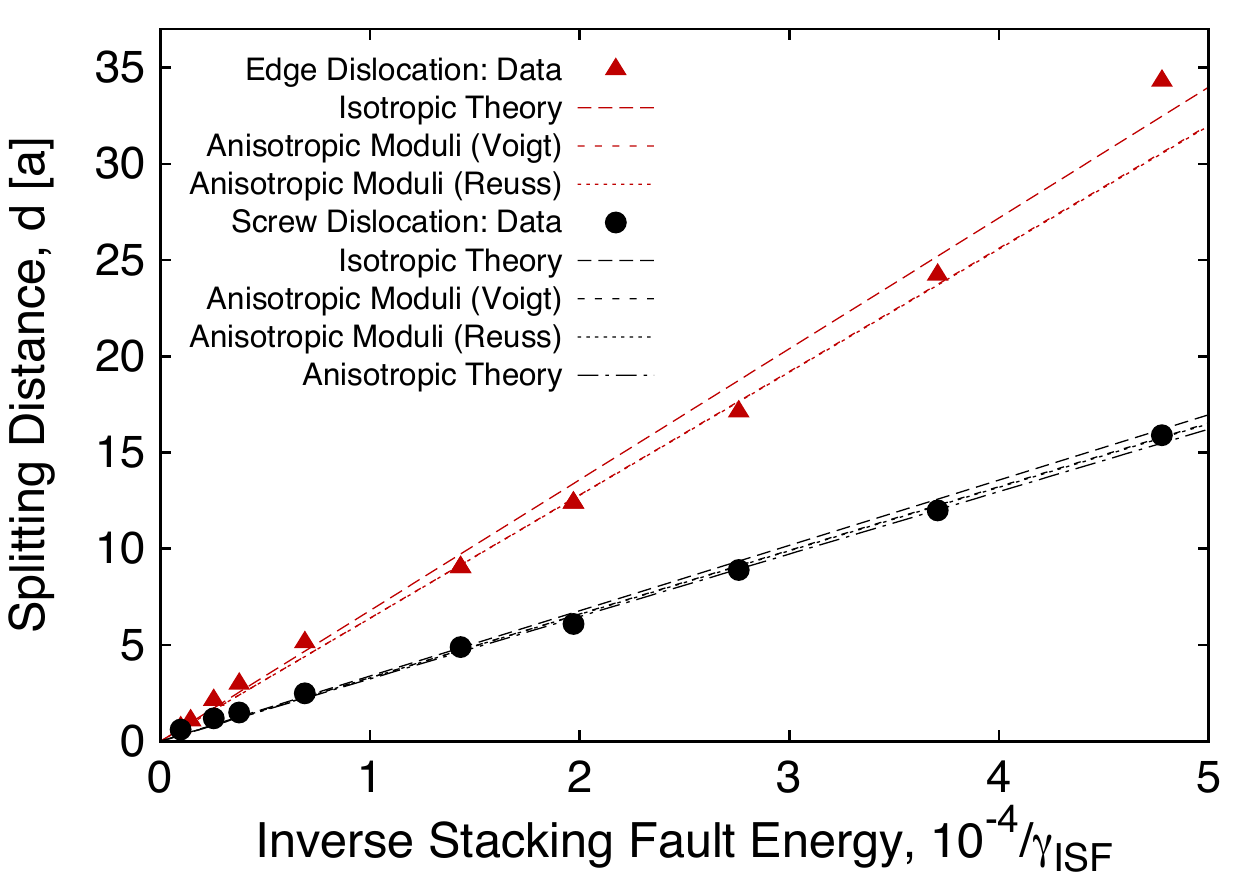} 
 }
 \vspace{-.25cm}
\caption[]
{\label{dvsgamma}
(Color online)
Equilibrium splitting distance between Shockley partials within
dissociated perfect $\vec{b}=a/2[110]$ edge and screw dislocations,
shown as a function of $\gamma_{\rm ISF}^{-1}$.
Lines represent the predictions of isotropic and anisotropic 
continuum elastic theories, with no adjustable parameters.
Model parameters are the same as those of Fig.\ \ref{c2gamma}.}
\end{figure}

We have measured $d_{\rm eq}$ for both edge and screw dislocations in
the modulated PFC model, as a function of $\sfe$. 
Base parameter values 
$n_0=-0.48$, $r=-0.63$,
and $B^x=1$ were used,
giving an fcc lattice constant $a=1.8537$.
The crystal orientation of Section \ref{sec:models} was employed
(see Fig.\ \ref{edgescrew}),
though in this case with periodicity broken in the $z$-direction by a
thin layer of liquid. A single perfect $\vec{b}=a/2[110]$ edge dislocation
with line direction $[\bar{1}12]$ was initialized at $(L_y/2,L_z/2)$ using
system dimensions $(L_x,L_y,L_z)=(16,3986,1820)$ and 
$\Delta x=a\sqrt{3/512}$ (203,668 atoms).
Perfect $\vec{b}=a/2[110]$ screw dislocation dipoles with $\langle 110\rangle$ 
line directions
were initialized similarly at $(L_x/4,L_z/2)$ and $(3L_x/4,L_z/2)$
with $(L_x,L_y,L_z)=(4278,10,1820)$ and
$\Delta x=a/\sqrt{200}$ (107,692 atoms).
The symmetry of the screw dislocation displacement field in the
$x$-direction necessitates a dipole configuration.
In both cases the diffusive dynamics of Eq.\ (\ref{eq:pfcdyn1}) were used.
Results are shown in
Fig.\ \ref{dvsgamma}, including comparisons with Eq. (\ref{deq}) and
the available anisotropic theories.

The expected linear trend is clearly observed for both dislocation types,
and the agreement with linear elastic predictions is excellent.
The fully isotropic theory very slightly overestimates $d_{\rm eq}$, 
while the isotropic theories
with approximate anisotropic moduli of Voigt and Reuss \cite{hirthlothe} 
and the fully anisotropic theory (which is tractable only for the
screw dislocation) match the data extremely well \cite{foot4}.
We note that it was necessary to simulate relatively large systems to 
eliminate significant boundary and image effects.
The periodic image forces nonetheless induce a small bias toward larger $d$
that grows with $d$ and appears to become appreciable for 
$d \gtrsim 25a$ at this system size.
Otherwise, the largest proportional disagreement is seen at small values of $d$
where the linear elastic predictions are likely somewhat inaccurate
due to core overlap effects.
We believe that these results provide an important
validation of fundamental defect energetics and interactions within
the PFC description.

\subsection{Dissociation width vs.\ applied (Escaig) strain}
\label{subsec:escaig}
When the glide-inducing shear 
strain $\epsilon_{zy}$
is applied to either of the dissociated dislocation lines shown in
Fig.\ \ref{edgescrew}, the two partials will glide together with
a nearly constant separation (except at large velocities as discussed below).
This is because the components of their respective Burgers vectors in
the shear direction $[110]$ have the same sense.
This is not the case for shear 
strain $\epsilon_{zx}$.
The two partials have components in this shear direction $[\bar{1}12]$ 
with opposite sense,
which produces glide forces in opposite directions
that extend the faulted region in between.
This particular strain orientation is sometimes referred to as the
Escaig strain.
The new equilibrium separation for fixed 
stress $\tau_{zx}$
can again be calculated within isotropic elasticity theory,
and for a general dislocation one obtains \cite{byun03}
\be
d=\frac{\mu b_p^2 f(\theta_1,\theta_2)}{\pi(2\sfe-\tau_{zx}b_p|
\sin{\theta_2}-\sin{\theta_1}|)}
\label{escaigdgeneral}
\ee
where
$f(\theta_1,\theta_2)=\cos{\theta_1}\cos{\theta_2}+\sin{\theta_1}\sin{\theta_2}/(1-\nu)$, 
and $\theta_1$, $\theta_2$ are the angles between the overall dislocation line 
direction and the Burgers vector directions of the leading and trailing 
partials, respectively.
For the case of a screw dislocation, 
where $\theta_1=-30^{\circ}$ and $\theta_2=30^{\circ}$,
Eq.\ (\ref{escaigdgeneral}) reduces to
\be
d=\frac{2-3\nu}{8\pi(1-\nu)}
\frac{\mu b_p^2}{\gamma_{\rm ISF}-\tau_{zx}b_p/2}.
\label{escaigd}
\ee

Shear strain $\epsilon_{zx}$ was applied to the screw dipole 
configurations described
in the previous subsection by adding a penalty function to the first few
crystalline surface layers near the liquid boundaries and translating
the penalty field at some constant rate $\dot{\epsilon}_{zx}$ to drive the
external deformation.
An affine shear deformation was also applied to the entire crystal 
at each time step, and the inertial equation of motion, 
Eq.\ (\ref{pfcinertia}), with $\alpha=1$ and $\beta=1/100$ was used.
These features together should ensure that the resolved shear strain at 
the dislocation remains as close to the applied shear strain as possible.
Selected fixed values of $\epsilon_{zx}$ were held periodically to
allow the system to fully relax to a steady configuration.
Though a reliable method of directly quantifying stress in our 
simulations is currently lacking, 
the assumption of linear elasticity will produce accurate stress-strain
conversions whenever plastic flow / stress dissipation is negligible
and strain is not exceedingly large.
We can confidently apply this assumption here 
($\tau_{zx}=C_{zx}\epsilon_{zx}$)
to obtain results in terms of stress, which can be directly compared
with Eq.\ (\ref{escaigd}).
Expected errors are $< 1-2\%$ for $\epsilon_{zx} \lesssim 0.05$.

\begin{figure}[btp]
 \centering{
 \includegraphics*[width=0.48\textwidth]{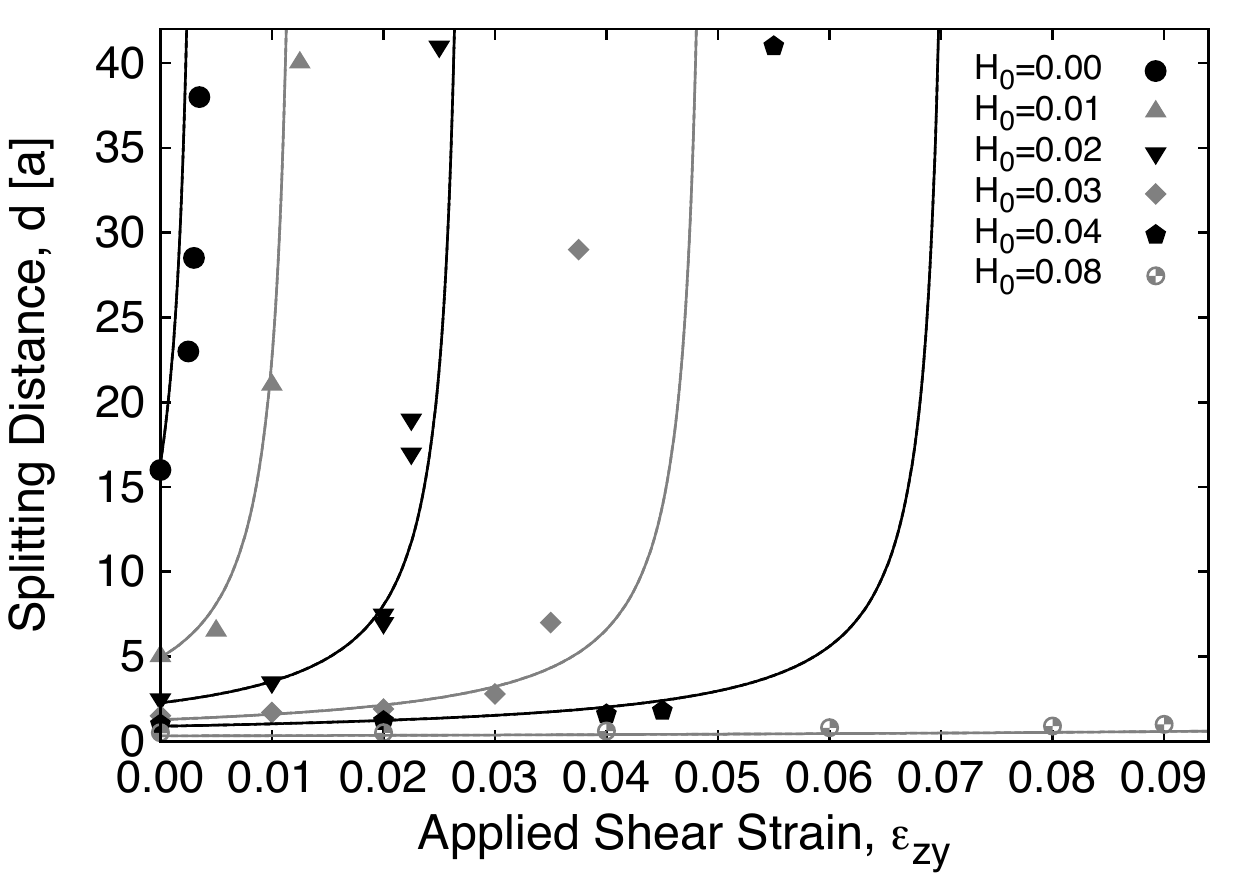} 
 }
 \vspace{-.25cm}
\caption[]
{\label{escaig}
Divergence of the
equilibrium splitting distance between Shockley partials within
dissociated perfect $\vec{b}=a/2[110]$ screw dislocations,
shown as a function of $\epsilon_{zx}$, the Escaig shear strain.
Lines represent the predictions of isotropic
continuum elastic theory, Eq.\ (\ref{escaigd}), as discussed in the text.
Model parameters are the same as those of Fig.\ \ref{c2gamma}.
}
\end{figure}

Simulation results are compared with the predictions of Eq.\ (\ref{escaigd}) 
in Fig.\ \ref{escaig}. 
The lines show Eq.\ (\ref{escaigd}) with elastic parameters as
determined from simulations.
The agreement is in general good for all values of $H_0$, with
the expected trend of diverging separation clearly visible. 
The strain at which divergence appears to occur agrees well with the predictions
for small $\sfe$, though not surprisingly shows increasing deviation as 
$\sfe$ and $\epsilon_{zy}$ become large.
These findings demonstrate that 
non-trivial defect properties associated with interactions between
defects and applied stresses can be accurately modeled with this approach.

\subsection{Peierls stress/strain for dislocation glide}
\label{subsec:peierls}
Dislocation motion is the dominant microscopic element of plastic deformation 
in most 
crystalline and polycrystalline metals.
The minimum strain required to move a dislocation, the Peierls strain
$\epsilon_{\rm P}$,
is therefore a fundamental material property that can qualitatively
alter a material's macroscopic plastic response, yield stress, etc.
fcc metals, in which dissociated dislocations and stacking faults are
prevalent, typically exhibit a characteristic slip activity that derives
directly from the nature of the dissociated dislocation structure
and the anisotropy of its Peierls barrier. 
Stacking faults are stable only on $\{111\}$ planes, and the 
Peierls barrier for motion of the partials is normally very small or negligible
within the same $\{111\}$ slip planes. The result is that a large
majority of slip activity occurs only within $\{111\}$ planes,
and a regime of qualitatively different plastic response emerges when
cross-slip between $\{111\}$ planes becomes active.
Thus a faithful description of the underlying Peierls barrier
is central to any model of crystal plasticity, and will naturally
give rise to many of the secondary features and processes that
emerge from the fundamental mechanism.

Accurate measures of Peierls stresses $\tau_{\rm P}$ in fcc metals 
have proven difficult to obtain through atomistic computations,
due to the smallness of $\tau_{\rm P}$ relative to typical
boundary image
stresses in finite size simulations \cite{mdpeierls2001}.
Typical estimates thus vary greatly in the literature, but
the most reliable values measured from molecular dynamics (MD) simulations
are believed to be on the order of $10^{-5}$-$10^{-4}\mu$ 
for fcc edge dislocations
and $10^{-4}$-$10^{-3}\mu$ for fcc screw dislocations
\cite{mdpeierls2001,hirthlothe,duesbery98}.
Experimentally determined values also vary greatly,
from $\sim10^{-6}-10^{-3}\mu$, depending on the method employed
\cite{basinski59,bujard87,nabarro97}.
Nonetheless, within these general ranges,
screw lines are expected to have larger barriers than edge lines,
and unsplit perfect dislocations to have larger barriers than
split partial configurations.

The Peierls strains for glide of edge and screw dislocations,
dissociated and undissociated, were measured by applying shear
to each system as described in the previous subsection.
Eq.\ (\ref{pfcinertia}) with $\alpha=1$ and $\beta=1/100$
was again employed, though in this case
shear strain $\epsilon_{zy}$ was applied. 
This orientation, unlike $\epsilon_{zx}$,
produces uniform glide of both
dislocation types, with split partials gliding in the same direction.
$\epsilon_{\rm P}$ was defined as the magnitude of applied strain at 
the instant a given defect has glided a distance equal to its Burgers 
vector magnitude $b$.

\begin{figure}[btp]
 \centering{
 \includegraphics*[width=0.48\textwidth]{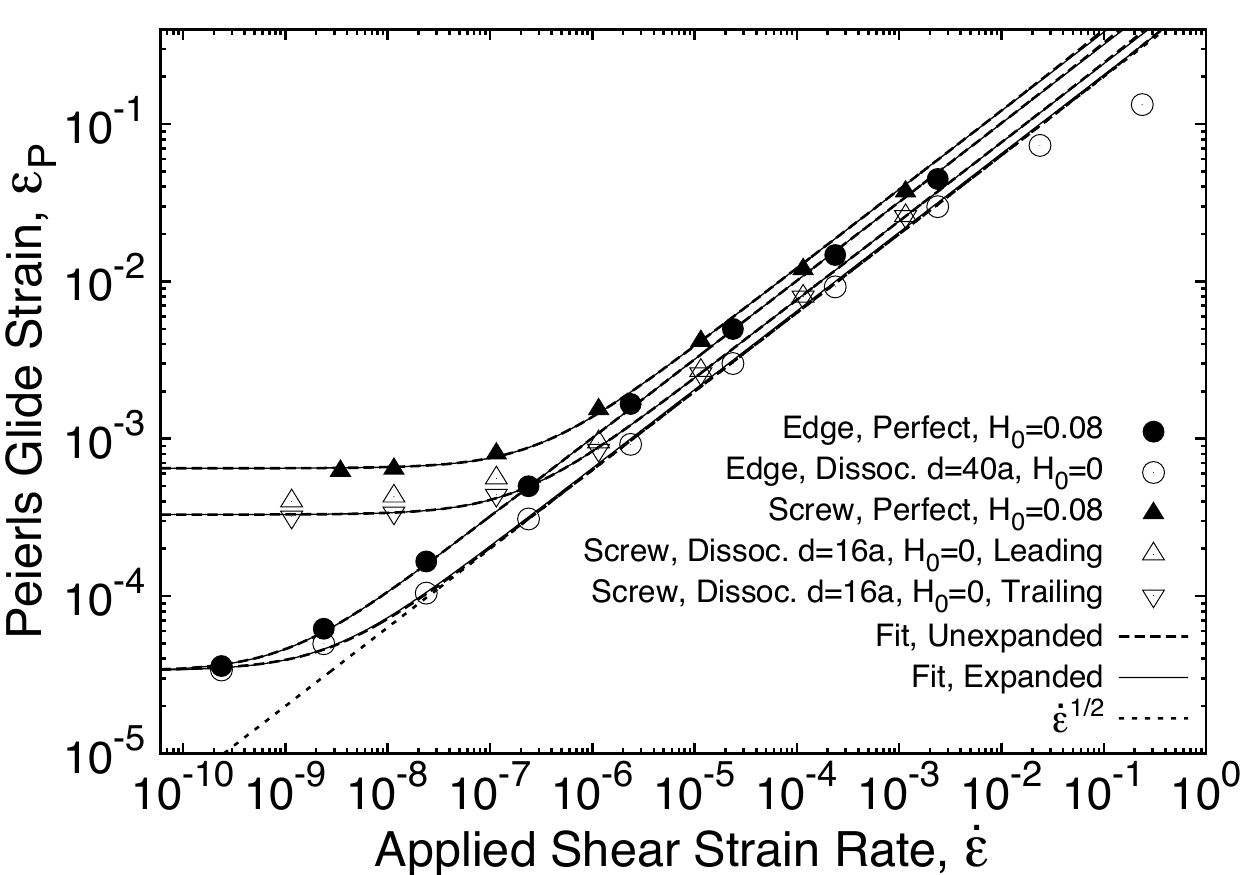} 
 }
 \vspace{-.25cm}
\caption[]
{\label{peierls}
Measured Peierls strain for glide of dissociated and undissociated
edge and screw dislocations, shown as a function of applied shear
strain rate.
The solid and dashed lines, which are indistinguishable, 
are fits to the functional forms presented in the text.
Model parameters are the same as those of Fig.\ \ref{c2gamma}.}
\end{figure}

The simulation results shown in Fig.\ \ref{peierls} bear out all of the
expectations noted above with only relatively minor quantitative deviations.
All dislocations exhibit the same roughly $\sqrt{\dot{\epsilon}}$
barrier strain dependence for sufficiently large shear rates. 
The cause of this behavior
is discussed below, but our primary interest is in the 
limiting values obtained at small shear rates.
Since the amount of dislocation motion in these simulations is negligible,
we can again apply linear stress-strain relations to report Peierls
stresses rather than strains
($\tau_{\rm P} \simeq \mu\epsilon_{\rm P}$).
The unsplit screw Peierls stress approaches a value 
within the expected range $\tau_{\rm P} \simeq 6\times10^{-4} \mu$,
while the average Peierls stress for the split screw
is slightly lower, $\tau_{\rm P} \simeq 3.5\times10^{-4} \mu$.
The measured values for the leading and trailing screw partials differ
by a small amount.
This may be a consequence of weakly asymmetric periodic image forces 
generated by the simulation boundaries, or of
an initial $d$ value that is incommensurate
with the periodicity of the Peierls potential, 
as discussed in Ref.\ \onlinecite{nabarro97}.

The widely split edge dislocation exhibits much less deviation between leading
and trailing $\tau_{\rm P}$ values, though the apparent limiting value 
is not smaller than that of the perfect edge dislocation to within
expected error. Both configurations approach a value 
that appears to extrapolate to $\tau_{\rm P} \simeq 3\times10^{-5}\mu$,
also within the expected range for typical fcc edge dislocations.
It was observed that edge dislocations tend to glide with a relatively
uniform, continuous progression, while screws exhibit a more
pronounced stick-slip hopping motion with each unit of translation,
indicative of a larger Peierls barrier.
These results together demonstrate that the fundamental features of fcc
Peierls stresses are captured quite reasonably by this relatively simple 
PFC model.

The apparent relation $\epsilon_{\rm P} \sim \sqrt{\dot{\epsilon}}$ 
for larger shear rates 
can be obtained as follows.
Based on our results, there appears to be an intrinsic time scale $\tau$
associated
with relaxation of an entire defect structure, inluding the core 
region and the long-range displacement fields.
If shear is applied slowly relative to $\tau$, the core region and
the long-range displacements can evolve cooperatively such that
the inherent Peierls barrier of the given dislocated system,
$\delta \tilde{F}_{\rm P}$, is realized. If
$\dot{\epsilon}^{-1}$ 
is large relative to $\tau$, then
the short-range and long-range components cannot relax with
optimal synchronization. This introduces
nonequilibrium effects and alternate relaxation processes that lead to
measurement of some larger effective barrier 
$\delta\tilde{F}_{\rm Eff}=\delta\tilde{F}_{\rm P}+\delta\tilde{F}_{\rm NEQ}$.

In the PFC formulation, 
slow relaxations near equilibrium are diffusive 
whether one uses Eq.\ (\ref{eq:pfcdyn1}) or Eq.\ (\ref{pfcinertia}).
Thus the intrinsic defect time scale $\tau$ can be approximated as
an exponential relaxation time, and
neglecting other relaxations, $\tilde{F}_{\rm NEQ}$
should be, to a first approximation, an
exponential function of the dimensionless parameter $\tau\dot{\epsilon}$.
We therefore may write
$\delta \tilde{F}_{\rm NEQ} \sim  e^{\tau\dot{\epsilon}}-1$ or 
$\delta \tilde{F}_{\rm NEQ} \sim  \tau\dot{\epsilon}$ in the small 
$\tau\dot{\epsilon}$ limit.
Combining this relation with a linear elastic response,
$\delta \tilde{F}_{\rm Eff} \sim \epsilon^2$, we obtain
$\epsilon_{\rm P} \sim \sqrt{\delta \tilde{F}_{\rm Eff}} \sim 
\sqrt{\delta \tilde{F}_{\rm P} + c \tau\dot{\epsilon}}$ 
where $c$ is a constant.
This form and the corresponding unexpanded form both fit our results well,
indicating that the competition between rate of applied shear and rate
of diffusive strain field relaxation leads to an increased effective 
Peierls strain with relevant control parameter $\tau\dot{\epsilon}$. 
The observed system size dependence of $\epsilon_{\rm P}$
is also consistent with this expression.

\subsection{Dynamic contraction of dissociated dislocations}
\label{subsec:contraction}
\begin{figure}[btp]
 \centering{
 \includegraphics*[width=0.48\textwidth]{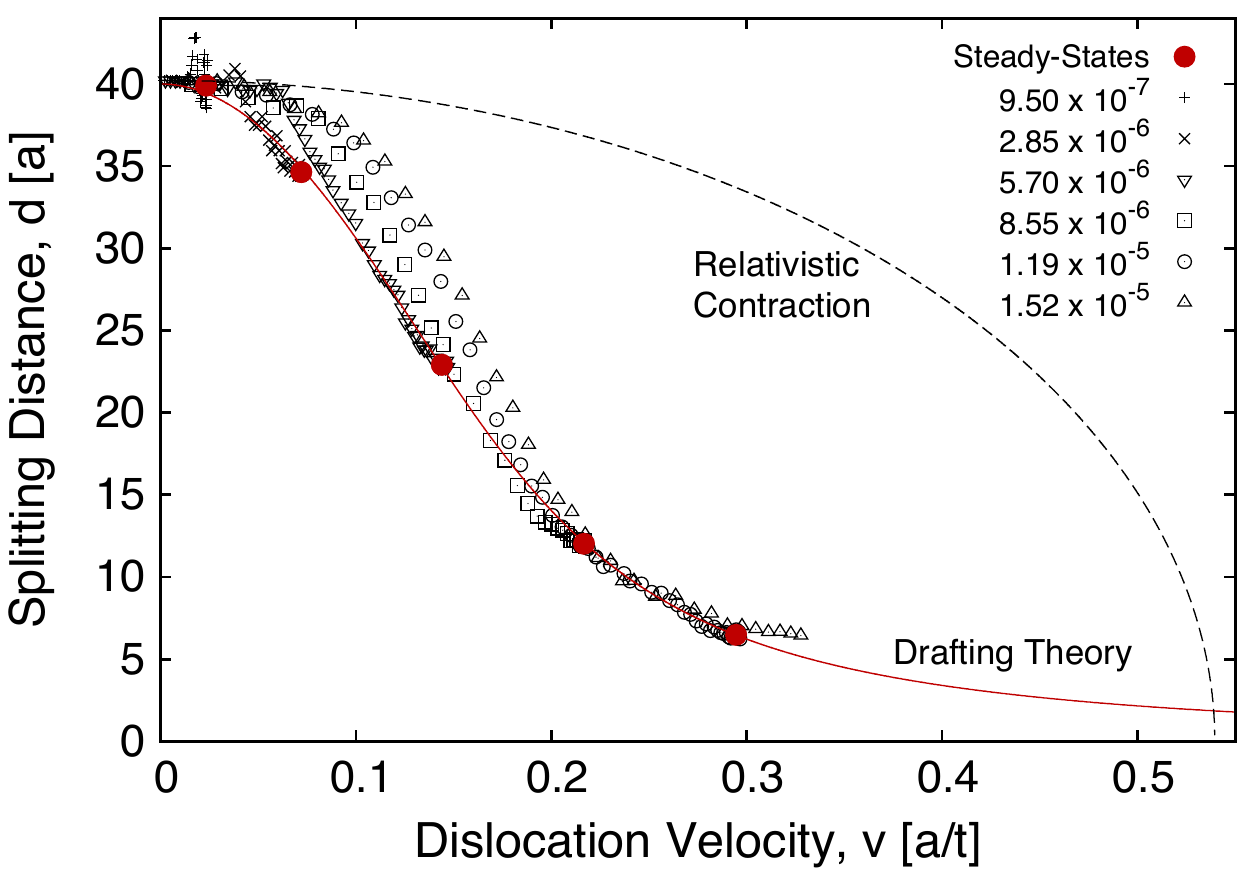} 
 }
 \vspace{-.25cm}
\caption[]
{\label{contraction}
(Color online)
Measured dissociation width of a gliding 
edge dislocation vs.\ dislocation velocity
for various applied shear strain rates.
Steady-state widths and velocities at each shear rate are separately indicated.
The solid line is a fit
to the functional form of the drafting theory presented in the text,
and the dashed line is an illustrative example of the behavior
expected from semi-relativistic effects 
($d \sim \sqrt{1-v^2/\alpha^2}$).
Model parameters are the same as those of Fig.\ \ref{c2gamma}.}
\end{figure}

When the glide-inducing stress $\tau_{zy}$ is applied to a dissociated
dislocation, the Peierls barrier is overcome quite quickly and 
glide is initiated within the slip system of the defect.
The structure of the extended defect may then be altered by dynamic
effects associated with navigation of the Peierls potential as well
as drag or damping forces of various origins.
Employing the dynamics of Eq.\ (\ref{pfcinertia}),
we applied fixed shear strain rates to the dissociated edge 
and screw
dislocation configurations already described, and allowed each
to approach a steady-state glide velocity $v$,
while montoring the separation between paired partials $d$.
Similar results were obtained for both dislocation types, 
and those for the $H_0=0$ edge dislocation are shown in Fig.\ \ref{contraction}.
For relatively low steady-state velocities ($v \lesssim 0.1\alpha$), 
$d$ was found to exhibit regular oscillations in time.
These oscillations, which can be seen in the
lowest shear rate data displayed in Fig.\ \ref{contraction},
are not unlike the so-called breathing modes
observed in MD simulations \cite{mdbreathing03,mdbreathing06},
though their period appears to be significantly longer and more regular
in these PFC simulations.

At higher velocities a dramatic decrease in $d$ was observed,
followed by a return approximately to the initial $d$ upon release of strain.
Since the shear strain $\epsilon_{zy}$ does not significantly alter $\sfe$
or the repulsive force between static paired partials, 
this contraction appears to be a fully dynamic effect.
We believe that the primary mechanism at work is a difference in the 
frictional drag force experienced by each partial at high PFC velocities
or, similarly, a reduction of the drag force exerted on the
extended defect structure as a whole.
Either scenario may result from the difference in screw-edge 
character between the two partials
\cite{hirthlothe} or may be driven by a 
cooperative drag-reduction
process analogous to
the phenomenon of drafting or slipstreaming in fluid dynamics.

At low velocities, where the effect of drag is small, 
$v$ is determined primarily by
the height of successive barriers in the Peierls potential.
Thus any forces that might change the dissociation width are small,
though they may become large enough to generate low-amplitude breathing modes.
As $v$ increases, the Peierls potential becomes less relevant and
the effect of frictional drag becomes large. 
In this regime, a dissociated defect
may experience a larger total drag force than an undissociated one 
due to its greater extension.
The total drag force may be reduced, thus increasing $v$ at a given strain,
by reducing $d$ such that the trailing partial assumes an optimized lower-drag
position within the leading wake or slipstream.
At large enough velocities, the reduction in drag force
experienced by the full defect under contraction may compensate
for the
increase in local defect energy. 
The energy of the entire system is reduced
by the excess strain relief
that is a consequence of faster dislocation glide.

We can modify the analysis given in Ref.\ \onlinecite{hirthlothe} to
incorporate such an effect into the predicted steady-state value of $d$.
The steady-state force balance between gliding paired
partials can be written\cite{hirthlothe}
\begin{eqnarray}
\sfe+B=D_1+\frac{A}{d}\nonumber\\
\sfe+D_2=C+\frac{A}{d}
\label{forcebalance}
\end{eqnarray}
where $A$ is an abbreviation of the terms multiplying $\sfe^{-1}$ 
in Eq.\ (\ref{deq}),
$B$ and $C$ are the forces per unit length 
on $\vec{b}_1$ and $\vec{b}_2$, respectively,
due to the applied shear, 
and $D_1$ and $D_2$ are the damping forces per unit length 
on the moving partials.
First we assume that both partials experience the same
force due to the applied shear strain ($B=C$) and that the nominal drag force
on each PFC dislocation obeys a Stokes' law form ($D_i \simeq C_d v$),
as shown in Ref.\ \onlinecite{pfcdisloc06}.
The drag coefficient should roughly be given by $C_d \simeq \mu/M$, where
$M = v/\epsilon \simeq 3$ is the partial dislocation mobility.
Finally, since we assume that $D_2$ decreases with $d$ ($D_2 = C_d v f(d)$)
a plausible functional form for $f(d)$ must be proposed.
We will use $f(d)=\delta d/(\delta d + c_v v)$ where
$\delta d=d+d_0$, and $d_0$, $c_v$ are constants.

A fit to the resulting equation for $d$ vs.\ $v$ is shown in Fig.\
\ref{contraction}. The only adjustable parameters in the fit are
$d_0$ and $c_v$, and the agreement is very good.
Semi-relativistic contraction of strain fields in the glide 
direction has also been predicted for dislocations moving at velocities near
the sound speed $\alpha$ of the crystal \cite{hirthlothe}. 
A curve illustrating this effect is shown in Fig.\
\ref{contraction}, where it is assumed that 
both partials experience the same contraction effect and reduce
$d$ accordingly.
This mechanism does not have a large effect until the sound speed is approached
and its form is not consistent with our results.
Furthermore, we observe similar behavior to that shown in Fig.\ 
\ref{contraction} when dynamics are given by Eq.\
(\ref{eq:pfcdyn1}), which does not introduce a sound speed.

The PFC description is generally best suited for examining behavior
under relatively low
strain rates or driving forces, where dislocation velocities tend not to
closely approach the sound speed.
Nonetheless, a better understanding of the nature of the
interaction between moving dislocations and the quasi-phonons described
by Eq.\ (\ref{pfcinertia}) would be useful in terms of
confirming that the artificially low sound speeds employed
in PFC simulations do not qualitatively 
alter the low and intermediate velocity dislocation dynamics.
Details of these issues are deferred to a future publication.

\section{Discussion and Conclusions}
\label{sec:conclusions}
The primary
factors controlling defect stability in PFC models have been examined, 
and it has been demonstrated that 
broad correlation kernels or elastically soft crystals
produce the greatest defect stability.
Maximally broadened kernels appear to be necessary for stabilization of certain 
defects such as planar faults
in the small wavenumber PFC approximation.
Higher wavenumber correlations 
with narrower peaks
were also shown to improve stability 
in some cases, but this feature leads to greatly reduced model efficiency.
The inherent conflict between crystal stability and defect
stability in broad kernel models
suggests potential limitations to the complexity of
structures that can be described by such models.
Any such limitations were shown to be non-factors in the case of the
primary defect structures in close-packed fcc crystals.

Stacking faults with low inherent stability were stabilized
in four PFC variants, indicating
that considerable defect stability can be obtained without losing crystal
stability.
The central defect structure in fcc plasticity, dissociated partial
dislocations with stacking faults, has been examined in some detail
and shown to be well-described by PFC methods.
The dependence of the equilibrium dissociation width on stacking fault 
energy has been shown to agree with continuum elastic predictions
for both edge and screw dislocations,
under zero and nonzero applied external stresses.
Peierls stresses for glide of both dislocation types have also been
measured and shown to fall within the typical ranges 
and relative magnitudes determined from
experiments and MD simulations.
Contraction of gliding pairs of partials has been observed at
high velocities and argued to be a consequence of large frictional
drag forces on widely split partials.

The findings presented in this article are intended to lay the groundwork for
larger-scale PFC studies of
plasticity in fcc crystals, with the potential to examine experimentally
relevant strain rates, 
nonconservative/climb-driven defect evolution,
and processes in which diffusing solute atoms interact with mobile and/or 
immobile defect structures.
These, in practice, are inaccessible
to conventional MD simulations.
Conversely, PFC models cannot reliably access the rapid time scales of MD and 
likely cannot predict some details of more complex defect atomic core
structures as accurately as MD.
Microscopic phase field (MPF) models of defects
\cite{PFDDreview2010}
are related to PFC in the sense that both employ phase field methodologies,
but MPF models utilize a top-down approach, explicitly building
the defect physics in by hand.
PFC models employ a far simpler, atomic-level free energy functional, 
from which all defect physics automatically emerge
with fewer imposed constraints and with
straightforward dynamical extensions.
Nonetheless,
by directly incorporating {\it ab initio} data such as
$\gamma$-surface energies \cite{PFgammasurf2004,PFgammasurf2011},
MPF models have proven highly effective in terms of predicting
static core-level defect structures and energies
with greater quantitative accuracy than current PFC descriptions.
At a more coarse-grained level,
both discrete dislocation dynamics models 
\cite{DDreview2009}
and coarse-grained phase field dislocation dynamics models
\cite{PFDDreview2010}
readily describe length scales inaccessible to MD, MPF, and PFC.
Similarly large length scales are in principle also accessible to
coarse-grained complex amplitude representations of PFC models
\cite{pfcRGnigel06a,pfcadmesh07,
pfcRGcoexist10,pfcbinaryamp10,pfcRGkarma10},
which still retain atomistic resolution.

Targeted PFC studies, examining more complex problems in 
defect physics that involve atomic length scales and diffusive time scales,
will be the subject of an upcoming publication.
Topics that have been analyzed and will be addressed include
climb fundamentals,
the structure of jogged dislocation lines, jog constriction,
jog pair annihilation, jog drag, screw dislocation cross-slip, 
formation and collapse of stacking fault tetrahedra from triangular Frank 
vacancy loops, Lomer-Cottrell or stair-rod dislocations,  
the interaction of dissociated edge and screw dislocations with
stacking fault tetrahedra, and dislocation creation mechanisms.

\begin{acknowledgments}
This work has been supported by the Natural Science and Engineering 
Research Council of Canada (NSERC), and access
to supercomputing resources has been provided by CLUMEQ/Compute Canada.
\end{acknowledgments}


\begin{thebibliography}{51}
\bibitem[{\citenamefont{foot1}(2012)}]{foot1}\citenamefont{
The latter implies that the Fourier transformed kernel $\hat{C}_2(k)$
is effectively shifted by a constant $u n_0^2$
relative to the kernel defined in Eq.\ (\ref{dftfree}).
To maintain consistency with existing notation
we will neglect this constant, and it will be implied that
$u n_0^2$ must be subtracted from
all kernels $\hat{C}_2(k)$ associated with Eq.\ (\ref{pfcfree}) 
if one wishes to recover the kernel that is fully consistent with 
Eq.\ (\ref{dftfree}).}
\bibitem[{\citenamefont{foot2}(2012)}]{foot2}\citenamefont{
The $\{111\}$ reflection will be broadened as well because the
three additional $\{111\}$ planes that are not parallel to the fault
will contain $1/(k_z-k_z^{(i)})^2$ modes.}
\bibitem[{\citenamefont{foot3}(2012)}]{foot3}\citenamefont{
Depending on local defect configurations
some dislocation and grain boundary structures can annihilate by glide alone,
leading to much lower energy barriers for removal. This effect is naturally
captured by CDFT and PFC descriptions.}
\bibitem[{\citenamefont{foot4}(2012)}]{foot4}\citenamefont{
The independent elastic constants were measured numerically
by using Eq.\ (\ref{eq:pfcdyn1}) 
and applying the appropriate strains, giving
$C_{11}=0.0967$, $C_{12}=0.0388$, and $C_{44}=\mu=0.032$.
An isotropic approximation of Poisson's ratio is therefore
$\nu=C_{12}/(C_{11}+C_{12})=0.2863$,
and the anisotropy ratio is $2C_{44}/(C_{11}-C_{12})=1.11$.
Thus there is a small degree of elastic anisotropy in this system,
as evidenced by the slightly lower slopes for the anisotropic predictions
in Fig.\ \ref{dvsgamma}.
The expressions for the Voigt and Reuss average anisotropic moduli
can be found in Ref.\ \onlinecite{hirthlothe}.
}
\end{thebibliography}
\end{document}